\documentstyle[prl,aps,epsfig]{revtex}

\newcommand{\beq}{\begin{equation}}
\newcommand{\eeq}{\end{equation}}

\begin{document}



\title{ 
       Fix-point Multiplier Distributions 
       in Discrete Turbulent Cascade Models
      }

\author{
B.\ Jouault$^{1}$,
M.\ Greiner$^{1}$,
and P.\ Lipa$^{2}$
}

\address{$^1$Max-Planck-Institut f\"ur Physik komplexer Systeme, 
             N\"othnitzer Str.\ 38, D--01187 Dresden, Germany }
\address{$^2$Dept.\ of Radiation Oncology, University of Arizona, 
             Tucson AZ-85721, USA }

\date{26.11.1998}

\maketitle

\begin{abstract}
One-point time-series measurements limit the observation of 
three-dimensional fully developed turbulence to one dimension. 
For one-dimensional models, like multiplicative branching processes,
this implies that the energy flux from large to small scales is not
conserved locally. This then
renders the random weights used in the cascade curdling to be different
from the multipliers obtained from a backward averaging procedure.
The resulting multiplier distributions become solutions of a fix-point
problem. With a further restoration of homogeneity, all observed
correlations between multipliers in the energy dissipation field can
be understood in terms of simple scale-invariant multiplicative
branching processes.
\end{abstract}

\vspace*{2cm}


\vspace*{4cm}

\noindent
CORRESPONDING AUTHOR: \\
Martin Greiner \\
Max-Planck-Institut f\"ur Physik komplexer Systeme \\
N\"othnitzer Str.\ 38 \\ 
D--01187 Dresden, Germany  \\
tel.: 49-351-871-1218 \\
fax:   49-351-871-1199 \\
email: greiner @ mpipks-dresden.mpg.de


\newpage

\section{Introduction}

As the classic of multi-scale processes, fully developed 
Navier-Stokes turbulence
remains to be a fortress, which besides many heroic attempts has not
surrendered to the cohorts of theoretical physicists \cite{FRI95}.
Hence, any progress in a thorough theoretical understanding should be
data driven. This means that for a characterisation and better
understanding of data there is first the need to develop simple
heuristic models and to try to learn the most from them. Only then,
with newly motivated and intriguing ideas and clever theoretical
technology, a Trojan horse can be found to take this fortress by storm.
--
This paper is not about the still to be found Trojan horse. It is about
the revelation of two pitfalls which appear during the early stage of
its attempted construction.

The energy dissipation field in fully developed turbulence reveals
striking, close to singular fluctuations. Empirically some features of
these intermittent fluctuations can be reproduced by 
multiplicative branching models \cite{MAN74,FRI78,SCH84,MEN87}.
Their particular strength lies in their simplicity and their inherent
self-similarity, which proofs useful when discussing multifractal
behaviour in the limit of a very large Reynolds number \cite{MEN91}.
These models rely on two assumptions:
{\bf (a1)}
the weight curdling, associated with the branchings and describing
the energy flux from large to small scales, is scale-independent, and
{\bf (a2)}
different branchings are independent of each other and are not
correlated. Of course, these assumptions have been directly tested 
with large Reynolds number atmospheric and wind channel data 
\cite{CHH92,SRE95,PED96,NEL96} and conclusions have been drawn that
{\bf (b1)}
assumption (a1) appears to hold whereas
{\bf (b2)}
assumption (a2) appears to be in conflict with the experimental
findings, where correlations between different, reconstructed
branchings have been observed. Although some modifications of the
cascade models have been proposed to include these effects \cite{SRE95},
many physicists think of it as a major drawback to the cascade models.

This criticism of the cascade models is by no means justified and this
brings us back to the pitfall-business mentioned in the beginning. The
immediate testing of the cascade model assumptions (a1) and (a2)
from data is not as straightforward as had been long anticipated. In
order to explain this point we have to understand how the measurements
are done:

An anemometer records one component, which is usually longitudinal, of the
velocity field as a time series in one point. Given that the fluctuation
strength is small this time series can be interpreted as a one-dimensional
spatial realisation of the velocity field at a fixed time; this is known
as Taylors frozen flow hypothesis. Since the energy dissipation tensor
is a functional of spatial derivatives of the velocity field, the spatial
realisation of the streamwise component of the velocity field can be
converted into a spatial realisation of the streamwise components of the
energy dissipation tensor. It is the statistical properties of this
one-dimensional energy dissipation field which are compared to, for
example, the multiplicative cascade models. The important thing to note
here is that the energy dissipation field is measured one-dimensional, 
but is generated from the three-dimensional dynamics of fully developed 
turbulence. If we assume a cascade dynamics to hold in three dimensions, 
then the energy flux from large to small scales
is conserved in every three-dimensional branching.
For a one-dimensional cut through the three-dimensional dynamics this,
of course, needs not to hold. As a consequence, branchings used in 
one-dimensional multiplicative cascade models should not conserve the
energy flux locally. Globally, that means on average, conservation of 
energy flux should be fulfilled. This then means, that the weights (or 
dynamical multipliers) used in the cascade evolution can not be reconstructed
from the one-dimensional cut through the energy dissipation field
resolved at the finest scale. With other words, the reconstructed 
multipliers obtained by smoothing from fine to large scales are not 
equal to the dynamical multipliers of the cascade evolution from large to 
fine scales.

A second pitfall occurring in the comparison of cascade models with data
has already been noticed previously \cite{GRE97}: multiplicative cascade
models lead to hierarchical structures which are not homogeneous. The 
time series measurements sketched above are unable to trigger on these
hierarchical structures. Hence, homogeneity has to be restored in the
cascade models first, before they are compared to data.

For a fair comparison between cascade models and data these two points
have to be taken into account:
{\bf (c1)}
no local conservation of energy flux in the branchings of one-dimensional
multiplicative cascade models since they represent one-dimensional
sections through an underlying three-dimensional cascade dynamics, and
{\bf (c2)}
restoration of homogeneity in the hierarchical multiplicative cascade 
models. Taking these two points into consideration we will show in this
Paper that multiplicative cascade models can be constructed, which
still respect assumptions (a1) and (a2) and which at the same time
are successful in explaining the presently available experimental
observations (b1) and (b2).

The organisation of the Paper is as follows: In Sect.\ II we summarise
the experimental facts about multiplicative cascade processes in fully
developed turbulence. Section III already reveals that three-dimensional
multiplicative cascade models, observed in one dimension,
lead to fix-point solutions for the various reconstructed
multiplier distributions which come close to the 
experimentally observed ones. 
Some analytical insight into the fix-point behaviour is given
in Sect.\ IV.
Sect.\ V considers three specific popular variants of one-dimensional
multiplicative cascade models, the asymmetric binomial,
the log-normal and the log-Poisson model,
and studies the implications of our two points (c1) and (c2). Conclusions
are given in Sect.\ VI.

\section{Experimental facts about turbulent cascade processes}

\subsection{Discrete multiplicative branching processes}

Multiplicative branching processes serve to model 
properties of the intermittent
fluctuations observed in the energy dissipation field of fully developed
turbulence. They relate the energy flux $E_L$ at some
integral scale $L$ to $E_r=E_L W_1(\lambda)\cdots W_J(\lambda)$ 
contained in a subinterval of size $r=\lambda^{-J}$ at scale $J$ 
by a product of mutually independent random weights $W_j(\lambda)$. 
The length ratio $\lambda$ represents a free parameter; 
the most simple and most commonly chosen value is given by $\lambda=2$. 
Adopting also this latter choice and dropping the dependence 
of the random weights $W_j$ on $\lambda$, the multiplicative branching 
processes are from now on denoted as binary multiplicative cascade
processes.

More precisely now, the energy flux $E_k^{(j)}$, contained in the
interval $0 \leq k < 2^j$ 
with length $r=L/2^j$ splits into a left (L) and right (R)
offspring interval, each of length $r/2$, whereby the content propagates
according to $E_{2k}^{(j+1)}=W_L E_k^{(j)}$ and $E_{2k+1}^{(j+1)}=W_R
E_k^{(j)}$, respectively.  For each breakup the two random weights 
$W_{L,R}\geq 0$ are chosen, independently from any
preceding breakup, according to a joint probability density $p(W_L,W_R)$.
The splitting function, as we denote the latter, guarantees a full
description of a binary breakup. For the
special case where the splitting function is concentrated along
the diagonal $W_L+W_R=1$, i.e.\ $p(W_L,W_R)=p(W_L)\delta(W_L+W_R-1)$, each
breakup strictly conserves energy flux. For all other forms the relation
$E_{2k}^{(j+1)}+E_{2k+1}^{(j+1)}=E_k^{(j)}$ needs not to hold; however, 
on average it should hold.

A particular simple example is the binomial splitting function
\begin{equation}
\label{21eins}
  p(W_L,W_R)  =  \frac{1}{2} 
                 \left[
                 \delta ( W_L - \frac{1+\alpha}{2} ) 
                 + \delta ( W_L - \frac{1-\alpha}{2} )
                 \right]
                 \delta (W_L+W_R-1)
                 \quad ,
\end{equation}         
which is known as the $p$-model \cite{MEN87}. Only two different local 
breakups are possible, each with the same probability. 
Here, energy flux is conserved for every local splitting.

In general, multiplicative cascade models are based on two assumptions: 
{\bf (a1)} 
the existence of a scale-independent splitting function $p(W_L,W_R)$ and 
{\bf (a2)}
statistical independence of the random weights $W_{L,R}$ at one breakup from
those of any other breakup.  
Once $p(W_L,W_R)$ is chosen,
the assumptions (a1) and (a2) allow to determine all moments and
scaling exponents of the energy dissipation $\varepsilon_r{=}E_r/r$ by inverse
Laplace transforms \cite{NOV71,GRE98}. Many experimental measurements of
moments and scaling exponents confirm that this simple construction reproduces
the measured multifractal aspects of the energy dissipation field amazingly
well (e.g. \cite{MEN91}).

\subsection{Experimental facts}

From a theoretical point of view the various parametrisations
suggested for the splitting function $p(W_L,W_R)$ serve to reproduce
the scaling exponents observed in the one-dimensional cuts of the 
energy dissipation field, the latter being converted from velocity
time series obtained from anemometers and employment of Taylors
frozen flow hypothesis. Within the experimental error bars, such
different models as the $p$-model \cite{MEN87}, 
the log-normal model \cite{OBU62,KOL62}, the log-Poisson model
\cite{SHE94,DUB94,SHE95} and others are all capable to describe the observed
scaling exponents. As a consequence of this model insensitivity some
efforts have been taken to extract information about the splitting
function directly from data \cite{CHH92,SRE95,PED96}. In the 
following we will briefly sketch the operational procedure and
summarise the essential results.

In the inertial range ($0 \leq j \leq J$) the energy dissipation
$\varepsilon (x,r)$
at length scale $r = L/2^j$ and position $x = k L/2^j$ is defined by
\begin{equation}
\label{22eins}
  \varepsilon_k^{(j)}
    =  \varepsilon \left( x = \frac{k L}{2^j} , r = \frac{L}{2^j} \right)
    =  \frac{1}{r} 
       \int_{x}^{x+r}
       \varepsilon ( x^\prime )  dx^\prime 
       \quad ,
\end{equation}         
where $0 \leq k < 2^j$. The experimentally measured `backward' energies 
then become
\begin{equation}
\label{22zwei}
  \bar{E}_{k}^{(j)}
    =  \frac{L}{2^j} \varepsilon_k^{(j)}
    =  \sum_{l=k 2^{J{-}j}}^{(k+1) 2^{J-j}-1}
       E_{l}^{(J)}
       \quad .
\end{equation}         
They are obtained by successive summations from the finest resolution 
scale $J$ to larger ones and are generally not equal to the `forward' 
energies $E_{k}^{(j)}$, which arise as intermediate states in 
the evolution of the cascade from larger to smaller scales.  For a 
clearer distinction we denote the former with a bar. 
--
Based on (\ref{22zwei}), so-called left, right and centred multipliers 
at scale $j$ and position $k$ are now operationally defined as
\begin{equation}
\label{22drei}
  M_{k,L}^{(j)}
    =  \frac{\bar{E}_{2k}^{(j+1)}} {\bar{E}_{k}^{(j)}}, 
       \quad 
  M_{k,R}^{(j)}
    =  \frac{\bar{E}_{2k+1}^{(j+1)}} {\bar{E}_{k}^{(j)}},
       \quad 
  M_{k,C}^{(j)}
    =  \frac{\bar{E}_{4k+1}^{(j+2)} + \bar{E}_{4k+2}^{(j+2)}} 
            {\bar{E}_{k}^{(j)}}.
\end{equation}         
They may be denoted as base 2 multipliers, because the associated
change of interval lengths refers to $\lambda=2$.
They depend on the relative position of parent to offspring intervals. 
Multiplier distributions $p(M^{(j)})$ are obtained by histogramming
all $0 \leq k < 2^j$ possible multipliers at a given scale and
averaging over many realisations.

For cascade models with an energy flux conserving splitting function
$p(W_L,W_R) = p(W_L) \delta(W_L+W_R-1)$ the random weights $W_{L/R}$ are
directly related to the multipliers $M_{L/R}$. Due to local energy flux
conservation we have 
$E_{2k}^{(j+1)} + E_{2k+1}^{(j+1)} 
 =  (W_{k,L}^{(j)}+W_{k,R}^{(j)}) E_{k}^{(j)}
 =  E_{k}^{(j)}
 =  \bar{E}_{k}^{(j)}$,
so that the `forward' energies $E_{k}^{(j)}$ are then identical to the
`backward' energies $\bar{E}_{k}^{(j)}$. Hence, 
$M_{k,L}^{(j)} = ( W_{k,L}^{(j)} E_{k}^{(j)} ) / E_{k}^{(j)}
               = W_{k,L}^{(j)}$,
and, analogously, $M_{k,R}^{(j)} = W_{k,R}^{(j)}$. For this case the
multiplier distributions $p(M_{L/R}^{(j)})$ reproduce the 
scale-invariant splitting function $p(W_L,W_R)$. 

Experimental analyses performed in atmospheric \cite{CHH92,SRE95} 
and wind tunnel turbulence \cite{PED96} have revealed that within 
the inertial regime the multiplier distributions
$p(M_{L/R}^{(j)})$ are to a very good degree independent of the length
scale. The following parametrisation has been found empirically
\cite{SRE95}:
\begin{equation}
\label{22vier}
  p(M_{L/R})
    =  \frac{\Gamma(2\beta)}{\Gamma(\beta)^2}
       \left[ M_{L/R} (1-M_{L/R}) \right]^{\beta-1}
       \quad ,
\end{equation}         
where $\Gamma$ stands for the gamma function and where 
$\beta \approx 3.2$. See Fig.\ 1. In order to link this scale-invariant
multiplier distribution directly to a splitting function, local
energy flux conservation in each binary breakup of the cascade process has to
be assumed. This then leads to
\begin{equation}
\label{22fuenf}
  p(W_L,W_R)
    =  \frac{\Gamma(2\beta)}{\Gamma(\beta)^2}
       \left[ W_L W_R \right]^{\beta-1}  
       \delta\left( W_L + W_R - 1 \right)
\end{equation}         
and the assumption (a1) of Sect.\ II.A of the existence of a 
scale-independent splitting function for multiplicative cascade models
appears to be confirmed. Notice, however, that the additional assumption 
of local energy flux conservation had to be employed to convert the
multiplier distribution (\ref{22vier}) into the splitting function
(\ref{22fuenf}).

What about assumption (a2) of Sect.\ II.A? 
If the multipliers $M_{k,L/R}^{(j)}$ are
independent from those at other scales and positions, as required
from the multiplicative cascade models, then the conditional multiplier
distributions
$p( M_{2k(+1),L/R}^{(j+1)} | M_{k,L/R}^{(j)} )$,
which correlate an offspring multiplier with its parent multiplier, 
should not depend on the latter, i.e.\
$p( M_{2k(+1),L/R}^{(j+1)} | M_{k,L/R}^{(j)} )
   =  p( M_{2k(+1),L/R}^{(j+1)} )$.
The experimental findings, however, indicate a dependency (see Fig.\ 1):
given a small parent multiplier 
$M_{k,L/R}^{(j)} \leq 1/2$
the conditional multiplier distribution
$p( M_{L/R}^{(j+1)} | M_{L/R}^{(j)} )$
is more narrow than the unconditional multiplier distribution
$p( M_{L/R}^{(j+1)} )$, 
and, given a large parent multiplier
$M_{k,L/R}^{(j)} \geq 1/2$
$p( M_{L/R}^{(j+1)} | M_{L/R}^{(j)} )$
becomes broader than 
$p( M_{L/R}^{(j+1)} )$;
here averages have been taken over all four possible combinations
$LL$, $LR$, $RL$, $RR$ of the parent and offspring multipliers. 
--
This observation has led to the conclusion about turbulent
cascade processes, that assumption (a1) appears to hold, whereas
assumption (a2) appears to be in conflict with the experimental
findings. To save some features of multiplicative cascade models
further extensions such as the correlated $p$-model \cite{SRE95}
have been suggested, which attempt to account for the effects seen in the
conditional multiplier distributions by introducing a dependency of the
local cascade branching on the magnitude of the local strain rate.

From an experimental point of view a multiplier analysis needs not only
be focused on left- (or right-) sided offsprings of a parent interval;
any relative position is qualified. Such a generalisation has been proposed
by Novikov \cite{NOV71}. In this context centred multipliers are
most often considered \cite{PED96,NEL96}; in our notation their 
definition is given in Eq.\ (\ref{22drei}). It has been observed that
the unconditional centred multiplier distribution $p(M_C^{(j)})$ is 
also scale-independent in the inertial regime, but is more narrow
than the unconditional $L/R$-multiplier distribution $p(M_{L/R}^{(j)})$;
see Fig.\ 2. The latter effect reflects the non-homogeneity of the breakup.
It can be explained, at least in a qualitative manner, within the binary
multiplicative cascade processes:
writing
$E_{4k+1}^{(j+2)} = W_{2k,R}^{(j+1)} E_{2k}^{(j+1)}
                  = W_{2k,R}^{(j+1)} W_{k,L}^{(j)} E_{k}^{(j)}$
and 
$E_{4k+2}^{(j+2)} = W_{2k+1,L}^{(j+1)} E_{2k+1}^{(j+1)}
                  = W_{2k+1,L}^{(j+1)} W_{k,R}^{(j)} E_{k}^{(j)}$,
respectively, and employing the energy flux-conserving splitting function
(\ref{22fuenf}), the distribution $p(M_{k,C}^{(j)})$ of Eq.\ (\ref{22drei})
becomes
\begin{eqnarray}
\label{22sechs}
  p(M_{k,C}^{(j)})
    &=& \int dW_{2k,L}^{(j+1)} dW_{2k,R}^{(j+1)} \,
              p(W_{2k,L}^{(j+1)},W_{2k,R}^{(j+1)})
         \int dW_{2k+1,L}^{(j+1)} dW_{2k+1,R}^{(j+1)} \,
              p(W_{2k+1,L}^{(j+1)},W_{2k+1,R}^{(j+1)})
         \nonumber \\
    & &  \int dW_{k,L}^{(j)} dW_{k,R}^{(j)} \,
              p(W_{k,L}^{(j)},W_{k,R}^{(j)})
         \;\,
         \delta\left(
              M_{k,C}^{(j)} 
              - \left[ 
                W_{2k,R}^{(j+1)} W_{k,L}^{(j)}
                + W_{2k+1,L}^{(j+1)} W_{k,R}^{(j)}
              \right]
         \right)
         \quad .
\end{eqnarray}         
The result is shown in Fig.\ 3.

In addition to the unconditional centred multiplier distributions also
a conditioning on centred parent multipliers has been investigated 
\cite{PED96}.
A small centred parent multiplier leads to a narrowing of the 
unconditional distribution; in addition the distribution is not anymore
symmetric and is skewed towards smaller multipliers. A large parent
multiplier, on the other hand, leads to a broadening of the distribution
with an asymmetry towards larger multipliers. Consult again Fig.\ 2.

\section{Three dimensional binomial cascade model as observed in a
         one-dimensional world}

The comparison between the multiplicative character of binary
branching processes and the experimental multiplier analysis is not
as straightforward and clear as it has been summarised in Sect.\ II.B. 
In fact, this comparison is not justified in a rigorous way as it
neglects two important features:
{\bf (c1)} violation of energy flux conservation in the bivariate 
splitting function and
{\bf (c2)} translational invariance. 
While the latter is the subject of the second Subsection, 
we now concentrate on the former.

\subsection{ Consequences of local non-conservation of energy flux }

In order to identify multipliers $M_{L/R}$ with weights $W_{L/R}$
of the bivariate splitting function, energy flux conservation ($W_L+W_R=1$)
of the latter had to be assumed. This assumption is not well justified.
The experimental analysis is restricted to measure the energy dissipation
field of three-dimensional turbulence only along a one-dimensional cut.
For the modelling this implies that, for example, if we were to consider
a three-dimensional multiplicative branching process with energy flux
conservation in every local three-dimensional breakup, 
then in one dimension the same process would appear as to violate local
conservation of energy flux.

Take, for example,
a three-dimensional binomial model, where a cube splits into eight subcubes
and where four randomly chosen subcubes get a large weight
$W_+ = (1+\alpha)/8$ and the other four get a small weight
$W_- = (1-\alpha)/8$. Along a one-dimensional cut this three-dimensional
branching is described by the bivariate splitting function
\begin{eqnarray}
\label{31eins}
  p(W_L,W_R)
    &=&  \frac{1}{2} \left\{
         \delta\left(  W_L - \frac{1+\alpha}{2}  \right)
         \left[
           \frac{3}{7}  \delta\left( W_R - \frac{1+\alpha}{2} \right)
           + \frac{4}{7}  \delta\left( W_R - \frac{1-\alpha}{2} \right)
         \right]
         \right.
         \nonumber \\
    & &  + \left.
         \delta\left(  W_L - \frac{1-\alpha}{2}  \right)
         \left[
           \frac{4}{7}  \delta\left( W_R - \frac{1+\alpha}{2} \right)
           + \frac{3}{7}  \delta\left( W_R - \frac{1-\alpha}{2} \right)
         \right]
         \right\}
         \nonumber \\
    &=&  p(W_L) \left[
         \frac{6}{7}\, p(W_R) + \frac{1}{7}\, \delta\left( W_L+W_R-1 \right)
         \right]
         \quad ,
\end{eqnarray}         
where
\begin{equation}
\label{31zwei}
  p(W_{L/R})  =  \frac{1}{2} \left[
                \delta\left( W_{L/R} - \frac{1+\alpha}{2} \right)
                + \delta\left( W_{L/R} - \frac{1-\alpha}{2} \right)
                \right]
                \quad .            
\end{equation}         
The second contribution in (\ref{31eins}) represents a small 
anticorrelation between weights $W_L$ and $W_R$ and is reminiscent of
energy-flux conservation in three dimensions. However, the uncorrelated,
first contribution in (\ref{31eins}) dominates and clearly does not
respect conservation of energy flux in one dimension.

Note, that for a one-dimensional binary multiplicative cascade process
with the splitting function (\ref{31eins}) the forward and backward 
energies $E_k^{(j)}$ and $\bar{E}_k^{(j)}$ are no longer identical. This 
is illustrated in Fig.\ 4, where the (forward) evolution of one cascade
realization down to scale $J$ is compared with the (backward) averaged
field. Only at the finest resolution 
scale the forward and backward fields are identical. It is due to the
non-conservation of energy flux in the splitting function that the two
fields are different for all other resolutions $0 \leq j < J$. Now,
the forward field is faithfully described by the splitting function
$p(W_L,W_R)$ depending on the left and right weight $W_L$ and $W_R$. 
For the backward field it is the left and right multipliers $M_L$ and
$M_R$ which are being deduced. Since the two fields are now different,
the multiplier distributions $p(M_{L/R})$ have no longer a simple 
relationship with the splitting function describing the physical
branching of the cascade; the former will be different from the latter.
This can be seen in Fig.\ 5a, where an averaging over $10^5$ simulated 
cascade realisations has been performed.
While the weights $W_{L/R}$ in the
forward evolution with the splitting function (\ref{31eins})
may take only two values $(1\pm\alpha)/2$,
the multipliers $M_{L/R}$ defined in Eq.\ 
(\ref{22drei}) take more and more distinct values as the
scale difference $J-j$ increases and becomes soon quasi-continuous.  Already
after three backward steps the left (and right) multiplier distributions
apparently converge to a scale-independent
limiting form which comes surprisingly close to a
parametrisation of the experimentally observed multiplier distribution
\cite{SRE95}, shown as continuous line.

Since the splitting function (\ref{31eins}) is symmetric under exchange
of $W_L \leftrightarrow W_R$, the unconditional 
right multiplier distributions are identical
to the left ones illustrated in Fig.\ 5a. Of course, this does not hold 
anymore for the centred multipliers $M_C$; their unconditional distributions
are depicted in Fig.\ 5b. As $J-j$ increases these distributions also
converge to a stable limiting form, which is now more narrow than the 
limiting form experienced for the unconditional left (or right) multiplier
distributions. At least, and this is very encouraging for the moment, the
trend is right when we think of the experimental facts presented in Sect.\
II.b.

The scale-independence of the limiting forms of the various unconditional
multiplier distributions suggest that they can be understood as solutions
of a fix-point problem. In Sect.\ IV we will further elaborate on this 
suggestion. For the moment we are further driven by curiosity and
ask whether we already observe some effects in conditional multiplier 
distributions. The simulations reveal that those themselves converge to
limiting forms after only three or four backward steps, starting from
the finest resolution scale $J$. Figures 6a and 6b show conditional
multiplier distributions at scale $j=3 < J=9$. The multiplier distribution
$p ( M_{L/R}^{(j)} | M_{L/R}^{(j-1)} )$
conditioned on the previous multiplier $M_{L/R}^{(j-1)}$,
where an average over all four possible daughter/parent combinations 
is performed, reveals a 
dependence on the latter. For a small parent multiplier, here
$0.2 \leq M_{L/R}^{(j-1)} \leq 0.4$, the distribution becomes somewhat broader
than the unconditional one, and for a large parent multiplier, here
$0.6 \leq M_{L/R}^{(j-1)} \leq 0.8$, the conditional distribution becomes a 
little more narrow. Although, in view of the experimental findings, 
these effects are in the right direction, we resist not to overstress them,
since the associated magnitude in the deviations from the unconditional
distributions appears still too small. A similar result is obtained for the
centred multiplier distribution 
$p ( M_C^{(j)} | M_C^{(j-1)} )$
conditioned on the parent multiplier $M_C^{(j-1)}$. For $M_C^{(j-1)}$ small,
here $0 \leq M_C^{(j-1)} \leq 0.5$, the distribution broadens, and for a 
large parent multiplier, here $0.5 \leq M_C^{(j-1)} \leq 1$, it narrows to 
some extend; note, however, that the experimentally observed shifts
to the right and left, respectively, are missing.

\subsection{Consequences of translational invariance}

Due to their hierarchical structure discrete cascade processes are 
obviously non-homogeneous. $n$-point correlation functions of the
generated energy dissipation field are not translationally invariant;
see Ref.\ \cite{GRE97} for a visualisation. This is in conflict with
the experimental findings, where $n$-point correlation functions are
deduced from velocity time series obtained from anemometers and 
employment of Taylors frozen flow hypothesis. Experimental realisations
of the energy dissipation field are obtained by chopping, more or less
arbitrarily, strings with length approximately equal to the correlation 
length $L$ out
of the recorded time series. For cascade model realisations, however,
we have knowledge about where the realisation of length $L$ begins and
ends. Hence, in order not to compare ``apples with oranges'', we have to 
introduce random translations within model realisations. Evidently, 
this will also influence multiplier distributions.

Two slightly different schemes have been used so far to restore
homogeneity in the (discrete) cascade models. In Ref.\ \cite{GRE97} two
independent cascade realisations, each having been evolved $J$ cascade
steps, have been attached and an observation window, having the same 
length as one cascade realisation, has been randomly placed within the
two attached realisations. Sampling over many realisations and many random
shifts then leads to homogenous $n$-point correlation functions. Moreover,
it has been found that
the multiplier distribution (\ref{22vier}) associated with the splitting
function (\ref{22fuenf}) remains invariant under this scheme. The same
findings also hold for a slightly modified scheme, which has 
been suggested in Ref.\ \cite{JOU98} and which we will employ in the 
following:
for the target resolution scale $J$ of a given integral length scale $L$
a longer cascade realization with $J+3$ steps is generated (corresponding to
an integral length scale $8L$) and an observation window of size $L$ is
shifted randomly by $t$ bins within $8L$. In other words, only the $2^J$ bins
of the generated $E^{(J+3)}_{k'}$, which lie within the randomly placed
observation window are considered, giving a simulated and translated
$E^{(J)}_k=E^{(J+3)}_{k'-t}$, where $t$ is a uniformly distributed integer
within $[0,\,7{\cdot} 2^J{-}1]$.  Then the multipliers are determined again by
(\ref{22zwei}) and (\ref{22drei}) and sampled over a large number of 
configurations and random shifts $t$.
In the simulations the second scheme to restore homogeneity will be 
applied with $J+3=12$ cascade steps, and the distributions have been averaged 
over $10^5$ independent realizations.

The results for the various multiplier distributions, determined by
cascade simulations with the splitting function (\ref{31eins}) and with 
the adopted scheme to restore translational invariance, are presented in
Figs.\ 7 and 8. For intermediate scales, here $1 \leq j \leq 5$, the 
unconditional left multiplier distribution $p(M_L^{(j)})$ converges again
to a stable limiting form, which is now in nearly perfect agreement
with the experimentally deduced left multiplier distribution (\ref{22vier})
with $\beta=3.2$. Apparently, the applied scheme to restore homogeneity
acts as to narrow the distribution to some extend; compare Figs.\ 5a and 7a.
Note a small detail: for very small scales, here $j=0$, the actual
multiplier distribution starts to deviate again from the limiting form
experienced for intermediate scales; this narrowing effect has already been
pointed out in Ref.\ \cite{GRE97}, where a slightly different scheme
to restore translational invariance had been applied to a binary cascade
with the splitting function (\ref{22fuenf}). The differences between
conditional left/right multiplier distributions, which are conditioned
on different values of
parent multipliers, are about the same with and without the 
proposed scheme to restore translational invariance; confer Figs.\ 6a and 
8a. The unconditional centred multiplier distributions also show
convergence to a slightly modified limiting form, but instead of a further
narrowing they experience a small, but noticeable asymmetry towards
smaller multipliers; compare Figs.\ 5b and 7b. This is a consequence
of the associated conditional centred multiplier distributions (Fig.\ 8b):
for a small centred parent multiplier the conditional distribution narrows
as in Fig.\ 6b, but this time it is also skewed towards smaller centred
daughter multipliers; a large centred parent multiplier leads to a broadening
of the conditional multiplier distribution with a simultaneous shift towards
larger centred daughter multipliers.

Overall, the various multiplier distributions become more realistic
once restoration of translational invariance is accounted for in the 
specific discrete cascade model with splitting function (\ref{31eins}), 
which is motivated from a three-dimensional binomial cascade process as 
observed in one dimension. Even the effects seen in the conditional
multiplier distributions go, at least, in the right direction. Evidently,
a non-energy flux conserving splitting function and restoration of
homogeneity are two key inputs to explain the current experimentally 
observed multiplier phenomenology within the simple cascade models. A 
further key input, namely asymmetric splitting functions, will be discussed
in Sect.\ V. But before that, we intend to enlighten the revealed
fix-point behaviour in the multiplier distributions and discuss a cascade,
where the splitting function is given by a symmetric Beta-distribution.

\section{Turbulent Cascade Model (Study I): Beta weight distribution}

The one-dimensional splitting function (\ref{31eins}) of a three-dimensional
binomial cascade process contains an anticorrelation between the left
and right weight. It is the result of local conservation of energy flux
in three dimensions. However, this anticorrelation remains small and,
when compared to the leading term in the one-dimensional splitting function,
may well be neglected. The one-dimensional bivariate splitting function
then takes on a factorised form:
\begin{equation}
\label{4eins}
  p(W_L,W_R) = p(W_L) p(W_R)    \quad .
\end{equation}         
From now on, this means Sects.\ IV and V, we will only discuss 
splitting functions belonging to this class. In this Section we begin with 
a Beta-distribution for $p(W_{L/R})$:
\begin{equation}
\label{4zwei}
  p(W_{L/R})
    =  \frac{\Gamma(2\beta)}{\Gamma(\beta)^2}
       \left[ W_{L/R} (1-W_{L/R}) \right]^{\beta-1}
       \quad .
\end{equation}         
One motivation for this distribution comes from the experimentally
observed left (or right) multiplier distribution, for which a 
Beta-distribution with $\beta=3.2$ represents a good parametrisation;
see discussion around Eq.\ (\ref{22vier}). Another motivation will be
outlined in Subsect.\ IV.b in connection with a first analytical insight
into the fix-point behaviour of multiplier distributions.

\subsection{Multiplier distributions without translational invariance}

Again, from simulations we extract the various multiplier distributions
belonging to a binary discrete cascade model with the factorised
splitting function (\ref{4zwei}). In this Subsection we do not employ
a scheme to restore translational invariance; this will be done in 
Subsect.\ IV.C. Both, the unconditional left as well as
centred multiplier distributions, illustrated in Fig.\ 9, converge to
respective limiting forms, where the latter distribution is more narrow
than the former. Only at the very finest scales, which means a large scale
index $j$, the distributions deviate from the limiting forms. At first
thought, it might look puzzling why the Beta-distribution ($\beta=3.2$), 
which is used for the splitting function, is not reproduced for all
scales in the unconditional $L/R$-multiplier distributions. Note, however,
that the used splitting function has a factorised form, so that there is
an enhanced probability that the two weights $W_L$ and $W_R$ are nearly
equal; since then $M_L \approx M_R \approx 0.5$ holds for the multipliers,
this explains the narrow $L/R$-multiplier distributions for $j$ close to 
$J$, the total number of performed cascade steps. If we had chosen a 
splitting function like (\ref{22fuenf}), reflecting a Beta-distribution
with local conservation of energy flux, then the extracted unconditional 
$L/R$-multiplier distributions would have been the same Beta-distribution
for all scales $j$. On second thought we should then ask, why does the
$L/R$-multiplier distribution converge back to the Beta-distribution? 
In the next Subsection we will work towards an answer of this question.

The conditional multiplier distributions
$p ( M_{L/R}^{(j)} | M_{L/R}^{(j-1)} )$ and 
$p ( M_{C}^{(j)} | M_{C}^{(j-1)} )$ also show convergence to limiting
forms. However, these limiting forms seem to be independent of the 
respective parent multiplier; this is illustrated in Fig.\ 10. This 
outcome differs from the results obtained in the previous Section. 
Whether effects in conditional multiplier distributions occur or not, 
apparently seems to depend on some specific features of the chosen
splitting function.

\subsection{Fix-point problem}

In the previous Subsection as well as previous Section we have started
to realize that various multiplier distributions, resulting from
one-dimensional cascade processes derespecting local conservation of
energy flux, appear to converge to scale-independent limiting forms.
These limiting multiplier distributions may well be understood as 
solutions of an associated fix-point problem. We will now formulate
this fix-point problem and discuss it for the particular case of a
factorised Beta-distribution splitting function.

We focus on the unconditional left multiplier distribution. We also do 
not consider the scheme to restore translational invariance, since it 
renders an analytical approach to be impracticable. According to 
Eq.\ (\ref{22drei}) the left multiplier $M_{k,L}^{(j)}$ at scale $j$
is defined as the ratio of two resumed energies $\bar{E}_{2k}^{(j+1)}$ 
and $\bar{E}_{k}^{(j)}$. The resumed energy $\bar{E}_{k}^{(j)}$ is equal
to the product of $j$ weights, drawn during the first $j$ cascade steps,
with the total resumed energy $\tilde{E}_{k}^{(J-j)}$ of a subcascade
with only $J-j$ cascade steps:
\begin{equation}
\label{42eins}
  \bar{E}_{k}^{(j)}
    =  W_{L/R}^{(0)} \cdots W_{L/R}^{(j-1)}
       \tilde{E}_{k}^{(J-j)}
       \quad .
\end{equation}         
For the other resumed energy we find
\begin{equation}
\label{42zwei}
  \bar{E}_{2k}^{(j+1)}
    =  W_{L/R}^{(0)} \cdots W_{L/R}^{(j-1)}
       W_{L}^{(j)}  \tilde{E}_{2k}^{(J-j-1)}
       \quad .
\end{equation}         
The first $j$ weights $W_{L/R}^{(0)}$, \ldots , 
$W_{L/R}^{(j-1)}$ in (\ref{42eins}) and (\ref{42zwei}) are identical.
The total resumed energy $\tilde{E}_{k}^{(J-j)}$ can be expressed as a
weighted sum of $\tilde{E}_{2k}^{(J-j-1)}$ of the left subbranch and
$\tilde{E}_{2k+1}^{(J-j-1)}$ of the right subbranch:
\begin{equation}
\label{42drei}
  \tilde{E}_{k}^{(J-j)}
    =  W_{L}^{(j)} \tilde{E}_{2k}^{(J-j-1)}
       + W_{R}^{(j)} \tilde{E}_{2k+1}^{(J-j-1)}
       \quad ;
\end{equation}         
this follows from (\ref{22zwei}). Note, that
$\tilde{E}_{L}^{(J-j-1)} = \tilde{E}_{2k}^{(J-j-1)}$ and
$\tilde{E}_{R}^{(J-j-1)} = \tilde{E}_{2k+1}^{(J-j-1)}$ are independent
variables. Combining (\ref{42eins}) -- (\ref{42drei}), the left
multiplier then becomes:
\begin{eqnarray}
\label{42vier}
  M_L^{(j)}
    &=&  \frac{ W_L^{(j)} \tilde{E}_{L}^{(J-j-1)} }
              { W_L^{(j)} \tilde{E}_{L}^{(J-j-1)} +
                W_R^{(j)} \tilde{E}_{R}^{(J-j-1)} }
         \nonumber \\
    &=&  \left(  1 + 
         \frac{ W_R^{(j)} }{ W_L^{(j)} }
         \frac{ \tilde{E}_{R}^{(J-j-1)} }{ \tilde{E}_{L}^{(J-j-1)} } 
         \right)^{-1}
         \quad .
\end{eqnarray}         
The quantities $W_L^{(j)}$, $W_R^{(j)}$, $\tilde{E}_{L}^{(J-j-1)}$,
$\tilde{E}_{R}^{(J-j-1)}$ are independent statistical variables; the first
two of them are drawn from a scale-independent splitting function
$p(W_L,W_R) = p(W_L) p(W_R)$. Since within the simulations the limiting form
of the left multiplier distribution appears to be scale-independent, 
this suggests that
the statistics of the resumed energies $\tilde{E}_{L/R}^{(J-j-1)}$
should also be scale-independent.

Denoting the probability density function of the resumed energy as 
$\tilde{p}(\tilde{E})$, we deduce from (\ref{42drei})
\begin{eqnarray}
\label{42fuenf}
  \tilde{p}(\tilde{E}^{(j^\prime)})
    &=&  \int {\rm d}W_L {\rm d}W_R p(W_L) p(W_R)
         \int {\rm d}\tilde{E}_L^{(j^\prime-1)} 
              {\rm d}\tilde{E}_R^{(j^\prime-1)}
              \tilde{p}(\tilde{E}_L^{(j^\prime-1)})
              \tilde{p}(\tilde{E}_R^{(j^\prime-1)})
         \nonumber \\
    & &  \delta \left(  
         \tilde{E}^{(j^\prime)} -
         \left[  W_L \tilde{E}_L^{(j^\prime-1)} +
                 W_R \tilde{E}_R^{(j^\prime-1)}  \right]
         \right)
         \quad ,
\end{eqnarray}         
where the substitution $j^\prime = J-j$ has been introduced. A 
Laplace transformation translates this equation into an equation for the
corresponding characteristic function:
\begin{equation}
\label{42sechs}
  Z^{(j^\prime)}[\lambda]
    =  \left\langle 
       \exp\left( \lambda\tilde{E}^{(j^\prime)} \right) 
       \right\rangle
       \quad = \quad
       \int \tilde{p}(\tilde{E}^{(j^\prime)}) 
            e^{\lambda\tilde{E}^{(j^\prime)}} 
            {\rm d}\tilde{E}^{(j^\prime)}
       \quad = \quad
       \left(
       \int {\rm d}W p(W) Z^{(j^\prime-1)}[W\lambda]
       \right)^2
       \quad .
\end{equation}         
If, here, we could find a scale-independent fix-point solution, then this
would explain the scale-independent limiting form of the left multiplier
distribution.

For the case of a splitting function with local conservation of energy
flux, i.e.\ 
$p(W_L,W_R) = p(W_L) \delta(W_L+W_R-1)$,
it is simple to find the fix-point solution. Eq.\ (\ref{42sechs}) then
generalises to
\begin{eqnarray}
\label{42sieben}
  Z^{(j^\prime)}[\lambda]
    &=&  \int  {\rm d}W_L  {\rm d}W_R \,  p(W_L,W_R) \,
         Z^{(j^\prime-1)}[W_L\lambda] \,  Z^{(j^\prime-1)}[W_R\lambda]
         \nonumber \\
    &=&  \int  {\rm d}W \,  p(W) \,
         Z^{(j^\prime-1)}[W\lambda] \,  Z^{(j^\prime-1)}[(1-W)\lambda]
         \quad ,
\end{eqnarray}         
which is fulfilled by 
$Z^{(j^\prime)}[\lambda] = Z[\lambda] = \exp\lambda$. 
Of course, then
$\tilde{p}(\tilde{E}) = \delta(\tilde{E}-1)$
and the left multiplier (\ref{42vier}) becomes $M_L = W_L$, so that its
distribution is scale-independent. We know this simple result already from
before, because for an energy flux conserving splitting function the 
left/right multiplier is always equal to the left/right weight.

For the case of a factorised splitting function, which of course does
not respect local conservation of energy flux, the fix-point solution
of (\ref{42sechs}) is much more difficult to find. In fact, for the moment
we are not aware of a general solution. We only know about a solution for
a specific choice of the splitting function \cite{BIA88}: the 
Beta-distribution given in Eq.\ (\ref{4zwei}). The fix-point characteristic
function $Z[\lambda]$ for the resumed energy is then the characteristic 
function of a Gamma-distribution:
\begin{equation}
\label{42acht}
  Z[\lambda]
    =  \left(  1 - \frac{\lambda}{2\beta}  \right)^{-2\beta}       
       \quad .
\end{equation}         
As has been demonstrated in Ref.\ \cite{BIA88}, this can be checked by
inserting (\ref{4zwei}) and (\ref{42acht}) into(\ref{42sechs}). Just to be
on the safe side, we have also performed a cascade simulation with 
$j^\prime \gg 1$
steps to determine the probability density function of 
$\tilde{E}^{(j^\prime)}$; the result is in perfect
agreement with the Gamma distribution
\begin{equation}
\label{42neun}
  \tilde{p}(\tilde{E})
    =  \frac{ (2\beta)^{2\beta} }{ \Gamma(2\beta) }
       \tilde{E}^{2\beta-1}  \exp\left( -2\beta\tilde{E} \right)
\end{equation}         
associated to the characteristic function (\ref{42acht}). 
--
According to (\ref{42vier}) the convolution of the Beta-distribution 
(\ref{4zwei}) for $W_L$ and $W_R$ with the Gamma-distribution (\ref{42neun})
for $\tilde{E}_L$ and $\tilde{E}_R$ leads to the limiting form of the left
multiplier distribution. As expected this convolution matches the 
limiting form (of Fig.\ 9) obtained from cascade simulations.

The fix-point distribution $\tilde{p}(\tilde{E})$ for the resumed energy
not only explains the scale-independent unconditional left (or right)
multiplier distribution, it also leads to scale-independent conditional
multiplier distributions. Changing notation as explained in Fig.\ 11, a
left daughter and parent multiplier become:
\begin{eqnarray}
\label{42zehn}
  M_{LL}  =  M_L^{(j)}
    &=&  \frac{  W_{LL} \tilde{E}_{LL}  }
              {  W_{LL} \tilde{E}_{LL} +
                 W_{LR} \tilde{E}_{LR}  }
         \nonumber \\
  M_{L}  =  M_L^{(j-1)}
    &=&  \frac{  W_L W_{LL} \tilde{E}_{LL} +
                 W_L W_{LR} \tilde{E}_{LR}  }
              {  W_L W_{LL} \tilde{E}_{LL} +
                 W_L W_{LR} \tilde{E}_{LR} +
                 W_R W_{RL} \tilde{E}_{RL} +
                 W_R W_{RR} \tilde{E}_{RR}  }
         \quad .
\end{eqnarray}         
All six weights $W_L$, $W_R$, $W_{LL}$, $W_{LR}$, $W_{RL}$, $W_{RR}$ 
as well as all four resumed energies
$\tilde{E}_{LL}$, $\tilde{E}_{LR}$, $\tilde{E}_{RL}$, $\tilde{E}_{RR}$
are independent variables and are drawn from scale-independent
probability density functions $p(W)$ and $\tilde{p}(\tilde{E})$, respectively. 
Hence, the probability density $p(M_{LL},M_L)$ is also scale-independent. 
Since the two multipliers $M_{LL}$ and $M_L$ partly consist of identical
random variables, they are correlated. This means that
$p(M_{LL},M_L) \neq p(M_{LL}) p(M_{L})$, which is responsible for the
correlation effects seen in the conditional left multiplier distribution
$p(M_{LL}|M_L)$; however, as has been realized in Fig.\ 10a, these effects
remain unnoticeably small for the special case of the 
Beta-distribution (\ref{4zwei}) entering into the splitting function
(\ref{4eins}). 
--
Note again, for an energy flux conserving splitting function we would always
have $M_L = W_L$ and $M_{LL} = W_{LL}$, so that
$p(M_{LL},M_L) = p(M_{LL}) p(M_{L})$ with no correlations.

With similar arguments the limiting forms of the unconditional as well as
conditional centred multiplier distributions can be traced back to the
fix-point probability distribution $\tilde{p}(\tilde{E})$ of the resumed
energy and the scale-independent splitting function $p(W_L,W_R)$. For 
example, in view of Fig.\ 11 the centred multiplier is expressed as
\begin{equation}
\label{42elf}
  M_{C} 
    =  \frac{  W_L W_{LR} \tilde{E}_{LR} +
               W_R W_{RL} \tilde{E}_{RL}  }
            {  W_L W_{LL} \tilde{E}_{LL} +
               W_L W_{LR} \tilde{E}_{LR} +
               W_R W_{RL} \tilde{E}_{RL} +
               W_R W_{RR} \tilde{E}_{RR}  }
         \quad ,
\end{equation}         
which clearly reflects a scale-independent distribution.

\subsection{Multiplier distributions with translational invariance}

Now, that we have gained some analytical insight into the occurrence of
scale-independent limiting
forms for the multiplier distributions, we again rely
on simulations and study the effect of restoring translational invariance
onto various multiplier distributions resulting from the factorised
splitting function (\ref{4eins}) with (\ref{4zwei}). As in Fig.\ 9, 
Fig.\ 12 illustrates the unconditional left and centred multiplier
distribution. The scale-independent limiting form of the unconditional
left multiplier distribution becomes again slightly more narrow after
application of the scheme to restore translational invariance, whereas
the corresponding centred multiplier distribution remains practically
unchanged.

The limiting forms of the conditional left/right multiplier distributions
show practically no dependence on the parent multiplier; only for extreme
small values of the latter some correlation effects are observed. See Fig.\
13 for an illustration. For the conditional centred multiplier distributions
the scheme to restore translational invariance has some more impact: the
daughter and parent multiplier become positively correlated, which means
that for large/small parent multipliers the occurrence of large/small 
daughter multipliers is enhanced. However, no broadening/narrowing of these
distributions is observed.

We conclude that a Beta-function parametrisation of the factorised
splitting function is not able to describe all effects observed in various
multiplier distributions. However, for this special case we succeeded to
understand the fix-point behaviour. In the next Section we will study some
more realistic splitting functions.

\section{Turbulent Cascade Model (Study II): 
         asymmetric weight distributions}

Adopting the factorised form $p(W_L,W_R) = p(W_L) p(W_R)$ for the 
splitting function there is more to expect for the weight distribution
$p(W_{L/R})$ from the one-dimensional section through a three-dimensional 
energy flux conserving cascade process. In an extreme 
three-dimensional branching event the parent energy is completely
transfered into one subcube; the other seven subcubes getting nothing.
The energy density contained in this one subcube is now eight times
as large as before, so that the energy contained in the projected
one-dimensional subinterval is four times as large as in the one-dimensional
parent interval. Hence, the values of the weights $W_{L/R}$ should not
be restricted to $0 \leq W_{L/R} \leq 1$, but to $0 \leq W_{L/R} \leq 4$.
Since the average weight will remain 
$\langle W_{L/R} \rangle = 1/2$, the weight distribution $p(W_{L/R})$
should then become asymmetric around $1/2$. 
--
In this Section three factorised splitting functions, having different
asymmetries, are picked and their implications on the various multiplier
distributions are discussed. We only show results obtained after 
application of the scheme to restore translational invariance.

\subsection{Asymmetric binomial weight distribution}

The simplest example of an asymmetric splitting function 
$p(W_L,W_R) = p(W_L) p(W_R)$ is a modification of (\ref{21eins}), 
(\ref{31eins}) and (\ref{31zwei}):
\begin{equation}
\label{51eins}
  p(W_{L/R})
    =  \frac{\alpha_2}{\alpha_1+\alpha_2}
       \delta\left(
         W_{L/R} - \frac{1-\alpha_1}{2}
       \right)
       +
       \frac{\alpha_1}{\alpha_1+\alpha_2}
       \delta\left(
         W_{L/R} - \frac{1+\alpha_2}{2}
       \right)
       \quad ,
\end{equation}         
and is known as the asymmetric $\alpha$-model \cite{SCH84}. For the
symmetric case, $\alpha_1 = \alpha_2$, it is identical to the splitting
function (\ref{31eins}), except for the missing small anticorrelation;
as a consequence the various multiplier distributions resulting from this
symmetric case are more or less  identical to those shown in Figs.\ 7 and 8.
For the asymmetric case we choose two different sets of parameters:
$\alpha_1 = 0.5$, $\alpha_2 = 0.3$ and $\alpha_1 = 0.3$, $\alpha_2 = 0.65$.
As depicted in Figs.\ 14a+b they represent two different asymmetries:
the former enhances weights, which are larger than $1/2$, whereas the latter
enhances weights, which are smaller than $1/2$.
The two sets of parameters have been adjusted in such a way that the 
limiting form of the unconditional left multiplier distribution comes
close to the experimentally observed parametrisation (\ref{22vier}). This
is illustrated in Figs.\ 15a+b. Also shown is the limiting form of the
unconditional centred multiplier distribution; both parameter 
cases lead to a narrow
distribution as central multiplier values around $1/2$ are more enhanced
than for left/right multipliers. Hence, as far as unconditional multiplier
distributions are concerned the two different asymmetric versions for the
splitting function lead to more or less indistinguishable results. 
For conditional multiplier distributions this is going to change!

The limiting form of the conditional $L/R$ multiplier distribution
practically shows no dependence on the parent multiplier once the first
set of parameters $\alpha_1=0.5$, $\alpha_2=0.3$ is employed. This is
exemplified in Fig.\ 15c. The second set of parameters 
$\alpha_1=0.3$, $\alpha_2=0.65$ reveals a different behaviour (see Fig.\ 15d):
for a large $L/R$-parent multiplier the conditional $L/R$ multiplier 
distribution is significantly broader than for a small $L/R$-parent 
multiplier; this trend, which would have also resulted without application
of the scheme to restore translational invariance,
is in agreement with experimental observations.
--
Also for the conditional centred multiplier distributions the two sets of
parameters $\alpha_1$, $\alpha_2$ lead to different results: for a 
large/small parent multiplier the conditional centred multiplier
distribution leads to a shift towards larger/smaller daughter multipliers.
Whereas this positive correlation is observed for both parameter sets
and is a consequence of the application of the scheme to restore
translational invariance alone, only
the second one leads to a further broadening/narrowing of the conditional
centred multiplier distribution, as observed from data.

These findings reveal that not every non-energy flux conserving splitting
function qualifies to reproduce the various multiplier distributions.
Moreover, in order to reproduce at least qualitatively the subtle effects
observed in the conditional multiplier distributions, a certain 
asymmetry should be included in the splitting function underlying the
discrete cascade process. This is in accordance with our intuitive picture
of a three-dimensional cascade process observed in a one-dimensional world.
In the next two Subsections we will test two popular models with
asymmetric splitting functions: the log-normal \cite{OBU62,KOL62} and the
log-Poisson model \cite{SHE94,DUB94,SHE95}.

\subsection{log-normal weight distribution}

For the log-normal model the weight distribution entering the factorised
splitting function is given by
\begin{equation}
\label{52eins}
  p(W_{L/R})
    =  \frac{2 e^{\sigma^2}}{\sqrt{2\pi}\sigma}
       \exp\left\{
         - \frac{1}{2\sigma^2}
         \left[
         \ln W_{L/R} + ( \ln 2 + \frac{3}{2}\sigma^2 )
         \right]^2
       \right\}
       \quad ,
\end{equation}         
where the normalisation is such that 
$\langle W_{L/R}^0 \rangle = 1$ and
$\langle W_{L/R} \rangle = 1/2$. It is a little unphysical because, in 
principle, values $W_{L/R} \geq 4$ can be drawn; however, in practice
the associated probability is negligible. The free parameter $\sigma$
is chosen such that the resulting limiting form of the unconditional
left multiplier distribution, obtained after application of the scheme
to restore translational invariance, comes close to the parametrisation
(\ref{22vier}); this gives $\sigma^2 = 0.45$. Consult Fig.\ 16a, where
also the resulting limiting form of the unconditional centred multiplier
distribution is depicted; again the latter is more narrow than the former.

Comparing the log-normal weight distribution (Fig.\ 14c) with the two
asymmetric $\alpha$-model weight distributions considered in the previous
Subsection (Figs.\ 14a+b), the asymmetry of the log-normal distribution
is more along the asymmetric $\alpha$-model distribution with the second
set of parameters $\alpha_1=0.3$, $\alpha_1=0.65$ than the other set.
If asymmetry towards smaller weights is indeed a criterion to reproduce
the various experimentally observed conditional multiplier distributions
on an at least qualitative level, then the log-normal model should also
qualify. Figs.\ 16b+c illustrate the resulting limiting forms of the 
various conditional multiplier distributions. The conditioning on a
left/right as well as centred parent multiplier leads to exactly the same
qualitative effects as we have already realized for the ``good''
asymmetric parametrisation of the $\alpha$-model.

\subsection{log-Poisson weight distribution}

With the assumption that the most intermittent structures in 
three-dimensional turbulence are one-dimensional isolated vortex lines
the scaling exponents
$\tau(n) = -c_1[n(1-c_2)-(1-c_2^n)]$
with $c_1=2$, $c_2=2/3$
have been derived for the energy dissipation field
$\langle (\epsilon_k^{(j)})^n \rangle = (L/2^j)^{\tau(n)}$
\cite{SHE94}. An inverse Laplace-transform directly leads to the weight
distribution
\begin{equation}
\label{53eins}
  p(W_{L/R})
    =  2^{-c_1}
       \sum_{m=0}^{\infty} 
       \frac{1}{m!} (c_1 ln2)^m 
       \delta\left(
       W_{L/R} - 2^{c_1(1-c_2)-1} c_2^m
       \right)
       \quad ,
\end{equation}         
which is of log-Poisson type \cite{DUB94,SHE95} and which we are now going 
to use as input into the bivariate factorised splitting function
$p(W_L,W_R) = p(W_L) p(W_R)$ for the evolution of a binary discrete
cascade process.

Observe, that there is no free parameter in the weight distribution 
(\ref{53eins}) as long as parameters $c_1$, $c_2$ are fixed by general
arguments. Hence, we can not finetune it to reproduce, for example,
the unconditional left multiplier distribution. The more remarkable it is,
that the respective limiting form obtained after application of the scheme
to restore translational invariance comes close to the parametrisation
(\ref{22vier}); see Fig.\ 17a. Also the limiting form of the unconditional
centred multiplier distribution is more narrow in comparison with the 
former distribution.

Comparing the weight distribution (\ref{53eins}) with the previously
discussed weight distributions, see Fig.\ 14, it appears that the 
log-Poisson distribution clearly favours weights larger than $1/2$ over
smaller values. In this respect it more or less resembles the
asymmetric binomial distribution with parameters $\alpha_1=0.5$, 
$\alpha_2=0.3$ and, hence, we expect no big effects in the conditional
multiplier distributions. As illustrated in Figs.\ 17b+c this is indeed
the case: the conditioning on a left/right parent multiplier leads only to
relatively small deviations from the respective unconditional distribution.
The conditional centred multiplier distributions again reveal a positive
correlation between daughter and parent multiplier, but no additional
broadening/narrowing is observed for a conditioning on large/small parent
multiplier values.

Instead of fixing the parameters $c_1$, $c_2$ by general arguments
\cite{SHE94}, they can also be used as free fit parameters. A comparison
of scaling exponents for longitudinal velocity structure functions obtained
from direct numerical simulation of low Reynolds number Navier-Stokes
turbulence \cite{GRO97} with those predicted from the log-Poisson
cascade model suggests the parameter choice $c_1=9.3$, 
$c_2=(0.947)^3=0.849$. With this choice the weight distribution 
(\ref{53eins}) becomes more symmetric around $W_{L/R}=1/2$ and, as a
consequence, the various unconditional and conditional multiplier
distributions nearly behave like those depicted in Figs.\ 7 and 8 for the
symmetric binomial model with splitting function (\ref{31eins}); the
multiplier correlations remain weaker than for the asymmetric binomial
model ($\alpha_1=0.3$, $\alpha_2=0.65$) or the log-normal model.

\section{Conclusions}

Fully developed Navier-Stokes turbulence is a three-dimensional nonlinear
process. With present experimental techniques it is not feasible to record
this 3+1-dimensional spatio-temporal dynamics directly; standard 
experimental observations record time series of the velocity field in one
point, which according to Taylors frozen flow hypothesis can be 
reinterpreted as a one-dimensional spatial cut through the 
three-dimensional velocity field at a given time. This reduction in
dimensions is an important point to keep in mind when analysing and
interpreting the ``one-dimensional'' data. -- 
In this Paper we have discussed the implications 
of this dimensional reduction on one specific 
inertial-range observable, the so-called multiplier distributions
associated to the energy dissipation field, by employing discrete
multiplicative branching processes. One-dimensional versions of the
latter should not locally conserve the energy flux from large to small
scales; moreover the splitting function should reflect an asymmetric
weight distribution. As a consequence of this, the dynamical forward
field used in the cascade evolution is not equal to the averaged
backward energy dissipation field. This leads to deviations from
perfect moment scaling \cite{GRE96}. Unconditional multiplier distributions 
obtained from the backward field do not match any longer the weight
distributions employed during the forward evolution, but converge to
scale-independent fix-point limiting forms which already come close to
the experimentally deduced multiplier distributions. An even better
agreement is obtained once also homogeneity is restored in the hierarchical
cascade processes. Then, not only unconditional but also conditional
multiplier distributions qualitatively match the experimentally observed
distributions. We conclude that the correlations observed in various
multiplier distributions are not in conflict with simple discrete
multiplicative branching processes, but are a consequence of a 
one-dimensional homogenous observation of a three-dimensional 
non-homogeneous and strictly selfsimilar cascade process.

So far the comparison between these simple cascade models and data
had only been qualitatively. One natural question to ask now is:
what is the correct, or at least best, one-dimensional splitting
function in accordance with data? This study has already shown, that
the asymmetric binomial model (with the second set of parameters) and
the log-normal model yield almost identical output for the various
multiplier distributions discussed so far. Hence, fine-tuning of 
different asymmetric parametrisations for the splitting function only
makes sense once further observables are included; these could be
for example multiplier or wavelet correlations \cite{GRE96,GRE95}
and $n$-point correlations of the energy dissipation field itself
\cite{GRE98}. Note, that in this last reference the inverse problem,
how to reconstruct the splitting function from $n$-point correlations,
has been solved in the pure discrete hierarchical cascade picture; 
however, the important aspect of restoring homogeneity destroys the
beautiful formalism. 
--
Only all these observables taken together are able to judge about the
relevance and qualities of (discrete) cascade and other empirical 
models!

\acknowledgements
One of us (B.J.) acknowledges support from the Alexander-von-Humboldt 
Stiftung. We thank J\"urgen Schmiegel for several fruitful discussions.

\newpage


\newpage
\widetext
\onecolumn

\begin{figure}
\begin{centering}
\epsfig{file=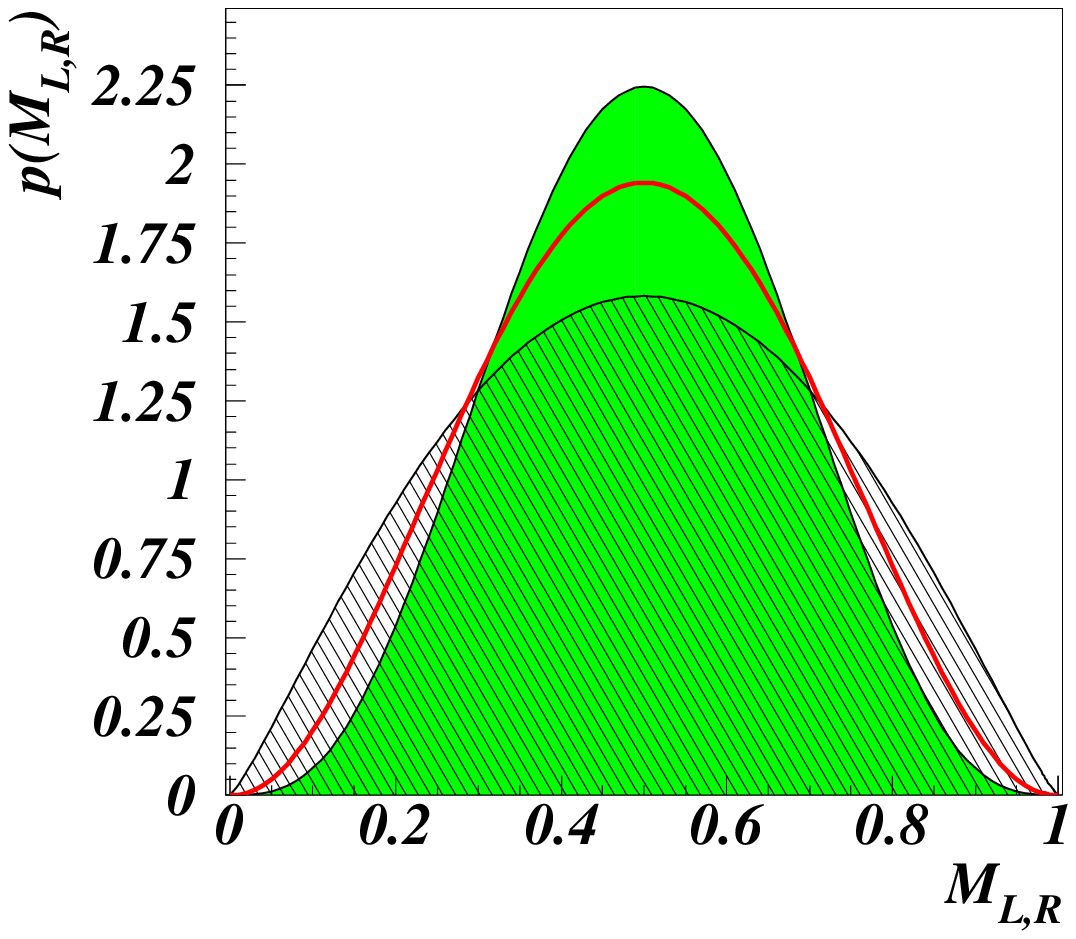,width=18cm}
\caption{
Schematic summary of the experimental findings with respect to the
scale-invariant $L/R$-multiplier distributions:
(thick solid line) unconditional distribution
$p( M_{L/R}^{(j+1)} )  
   =  p( M_{L/R}^{(j+1)} | 0 \leq M_{L/R}^{(j)} \leq 1 )$,
(grey) distribution
$p( M_{L/R}^{(j+1)} | 0 \leq \underline{M} \leq M_{L/R}^{(j)} 
                      \leq \overline{M} \leq 1/2 )$
conditioned on a small $L/R$-parent multiplier, and
(hatched) distribution
$p( M_{L/R}^{(j+1)} | 1/2 \leq \underline{M} \leq M_{L/R}^{(j)} 
                      \leq \overline{M} \leq 1 )$
conditioned on a large $L/R$-parent multiplier.
} 
\end{centering}
\end{figure}

\newpage
\begin{figure}
\begin{centering}
\epsfig{file=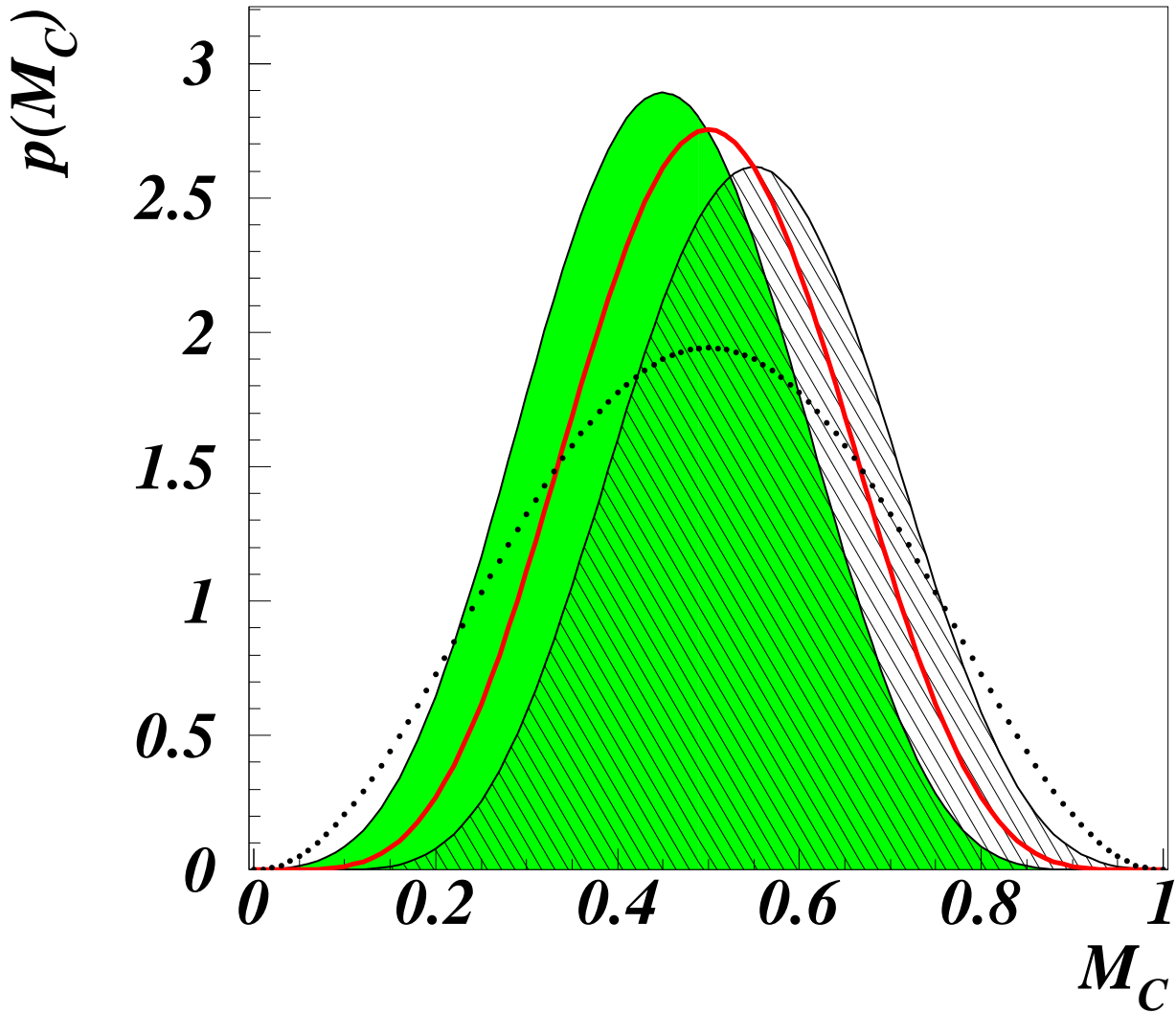,width=18cm}
\caption{
Schematic summary of the experimental findings with respect to the
scale-invariant centred multiplier distributions:
(thick solid line) unconditional distribution
$p( M_{C}^{(j+1)} )  
   =  p( M_{C}^{(j+1)} | 0 \leq M_{C}^{(j)} \leq 1 )$,
(grey) distribution
$p( M_{C}^{(j+1)} | 0 \leq \underline{M} \leq M_{C}^{(j)} 
                      \leq \overline{M} \leq 1/2 )$
conditioned on a small $C$-parent multiplier, and
(hatched) distribution
$p( M_{C}^{(j+1)} | 1/2 \leq \underline{M} \leq M_{C}^{(j)} 
                      \leq \overline{M} \leq 1 )$
conditioned on a large $C$-parent multiplier. 
For comparison, the unconditional $L/R$-multiplier distribution 
$p( M_{L/R}^{(j+1)} )$ of Fig.\ 1 is also shown as the 
thick dash-dotted curve.
} 
\end{centering}
\end{figure}

\newpage
\begin{figure}
\begin{centering}
\epsfig{file=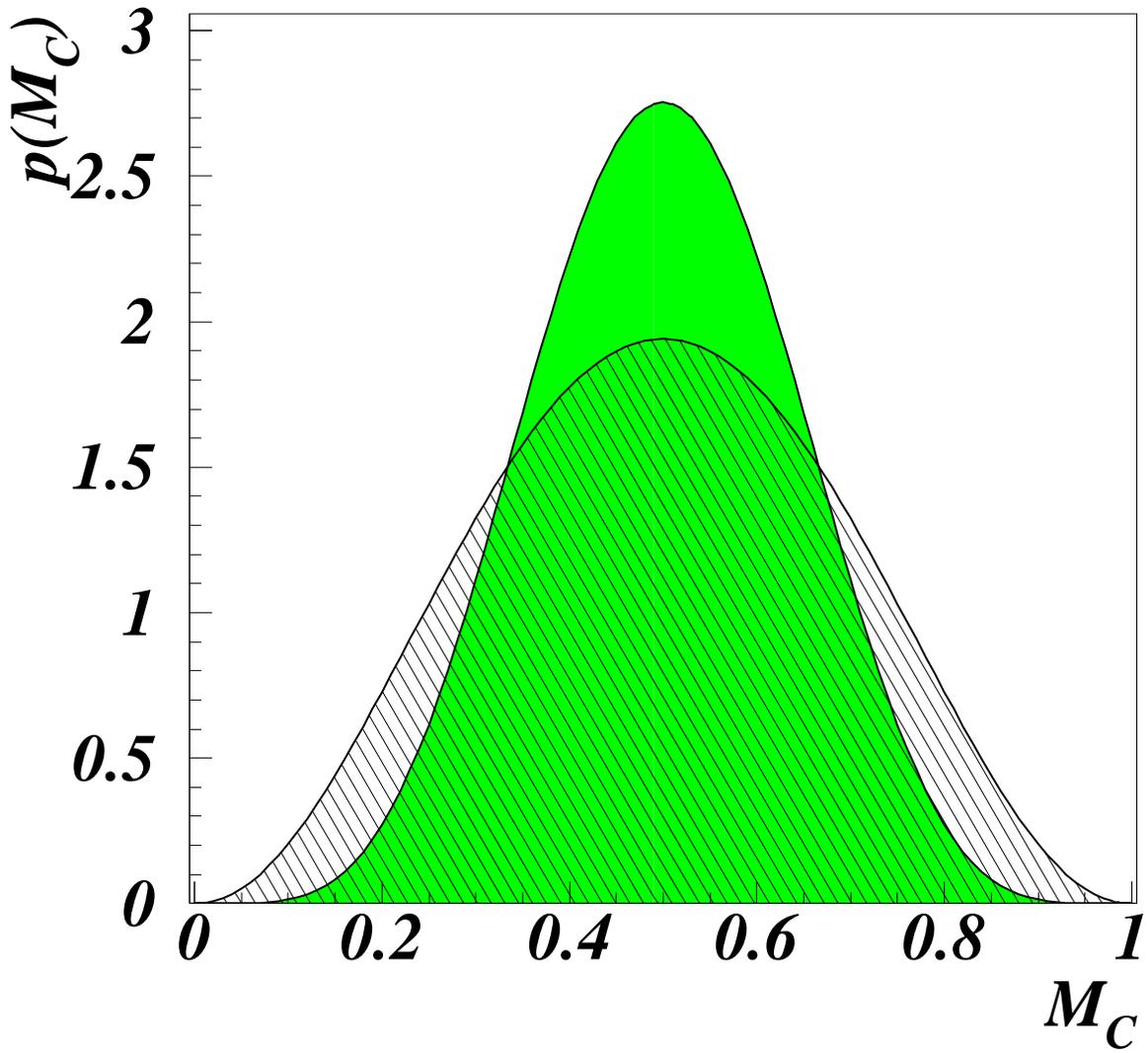,width=18cm}
\caption{
Unconditional distribution ({\protect\ref{22sechs}})
of centred multipliers $M_{C}^{(j)}$ (grey)
following from the $L/R$-splitting
function ({\protect\ref{22fuenf}}) (hatched).
} 
\end{centering}
\end{figure}

\newpage
\begin{figure}
\begin{centering}
\epsfig{file=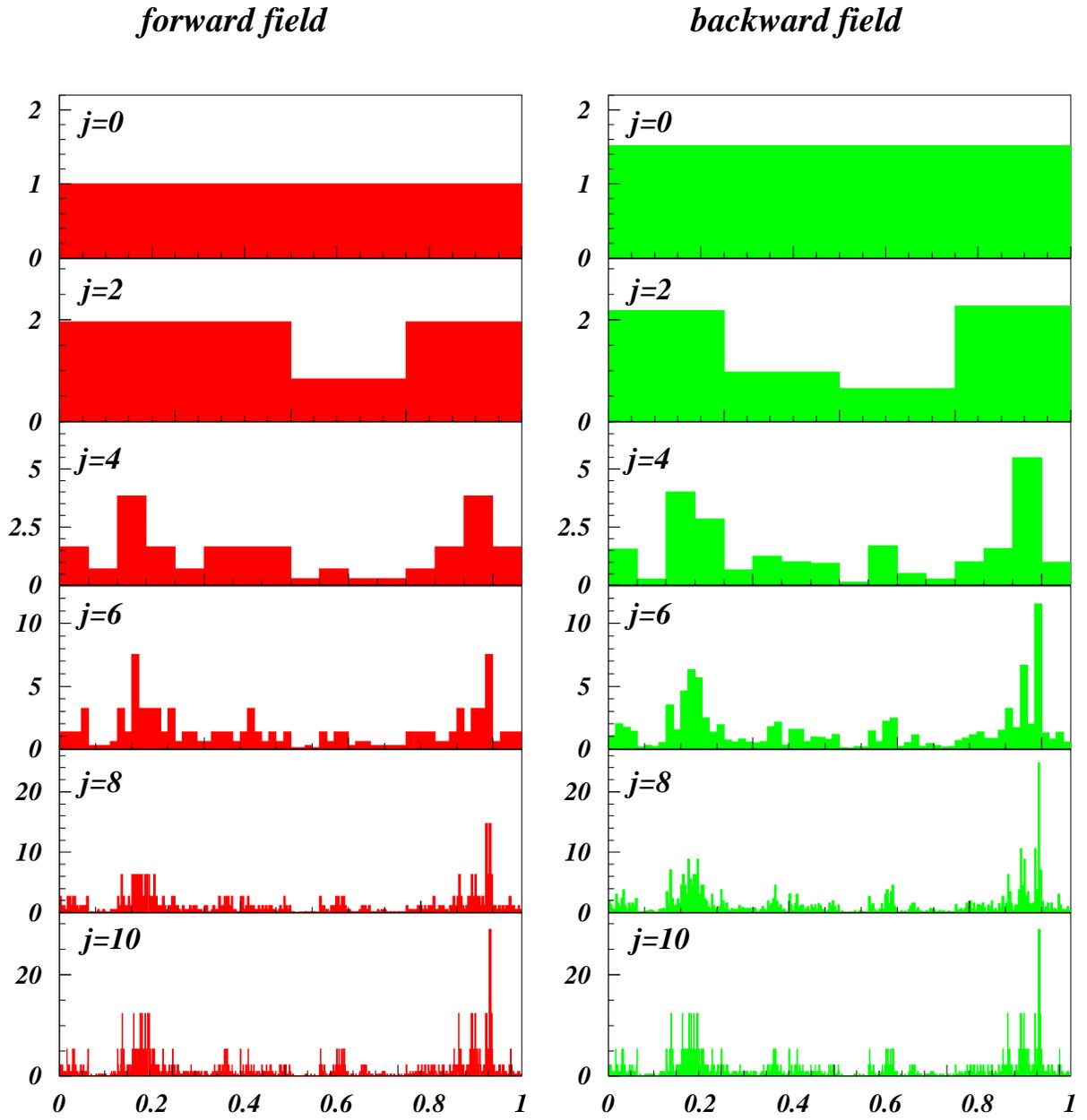,width=18cm}
\caption{
Cascade evolution of one energy dissipation field realisation 
from large to small
scales (left column) vs.\ integrated energy dissipation field from small
to large scales (right column). The non-energy flux conserving splitting 
function ({\protect\ref{31eins}}) has been employed with $\alpha=0.4$ and
$J=10$. Only every second scale $j = 0, 2, \ldots , 10$ is shown.
} 
\end{centering}
\end{figure}

\newpage
\begin{figure}
\begin{centering}
\epsfig{file=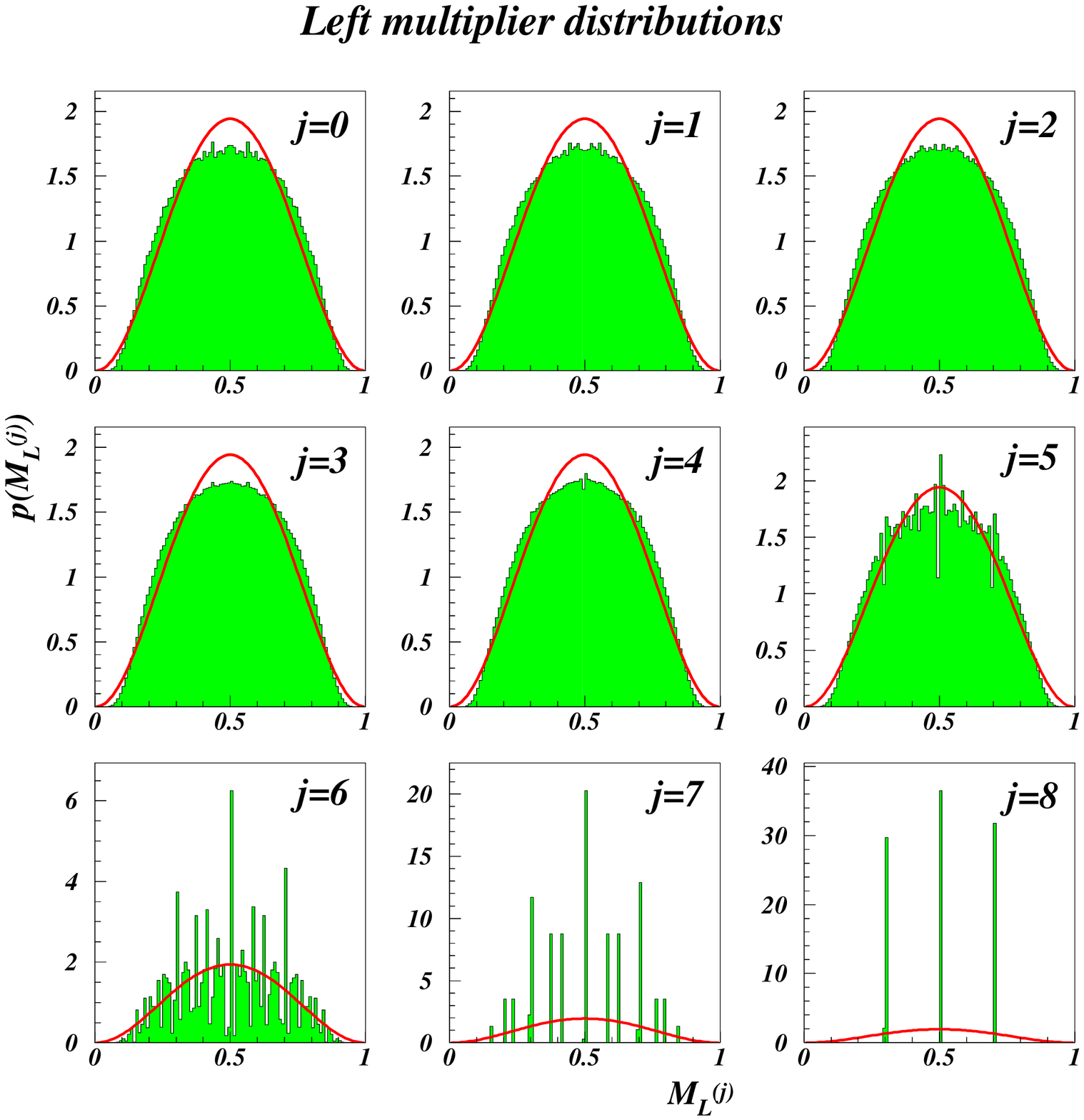,width=18cm}
\end{centering}
\end{figure}
\newpage
\begin{figure}
\begin{centering}
\epsfig{file=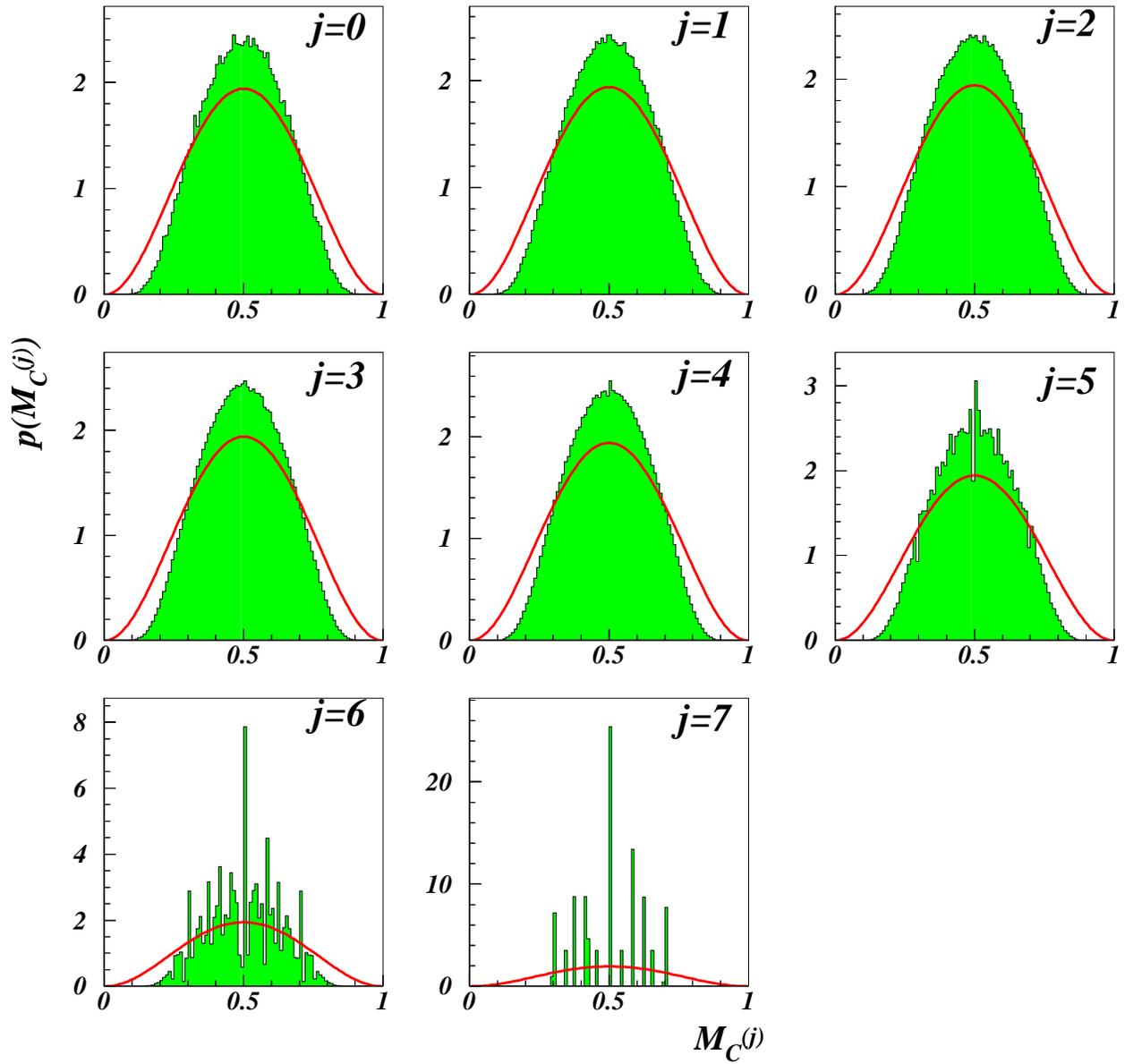,width=18cm}
\caption{
Convergence of the
(a) left and
(b) centred unconditional multiplier distributions 
$p(M_L^{(j)})$ and $p(M_C^{(j)})$ of the 
one-dimensionally observed three-dimensional binomial model
({\protect\ref{31eins}}) to a 
quasi-continuous limiting form.
Parameters are $J=9$, 
$\alpha=0.4$. For comparison the experimentally deduced 
Beta-function parametrisation ({\protect\ref{22vier}}) 
of the left (right) unconditional multiplier distribution with 
$\beta=3.2$ is shown as a solid line. 
} 
\end{centering}
\end{figure}

\newpage
\begin{figure}
\begin{centering}
\epsfig{file=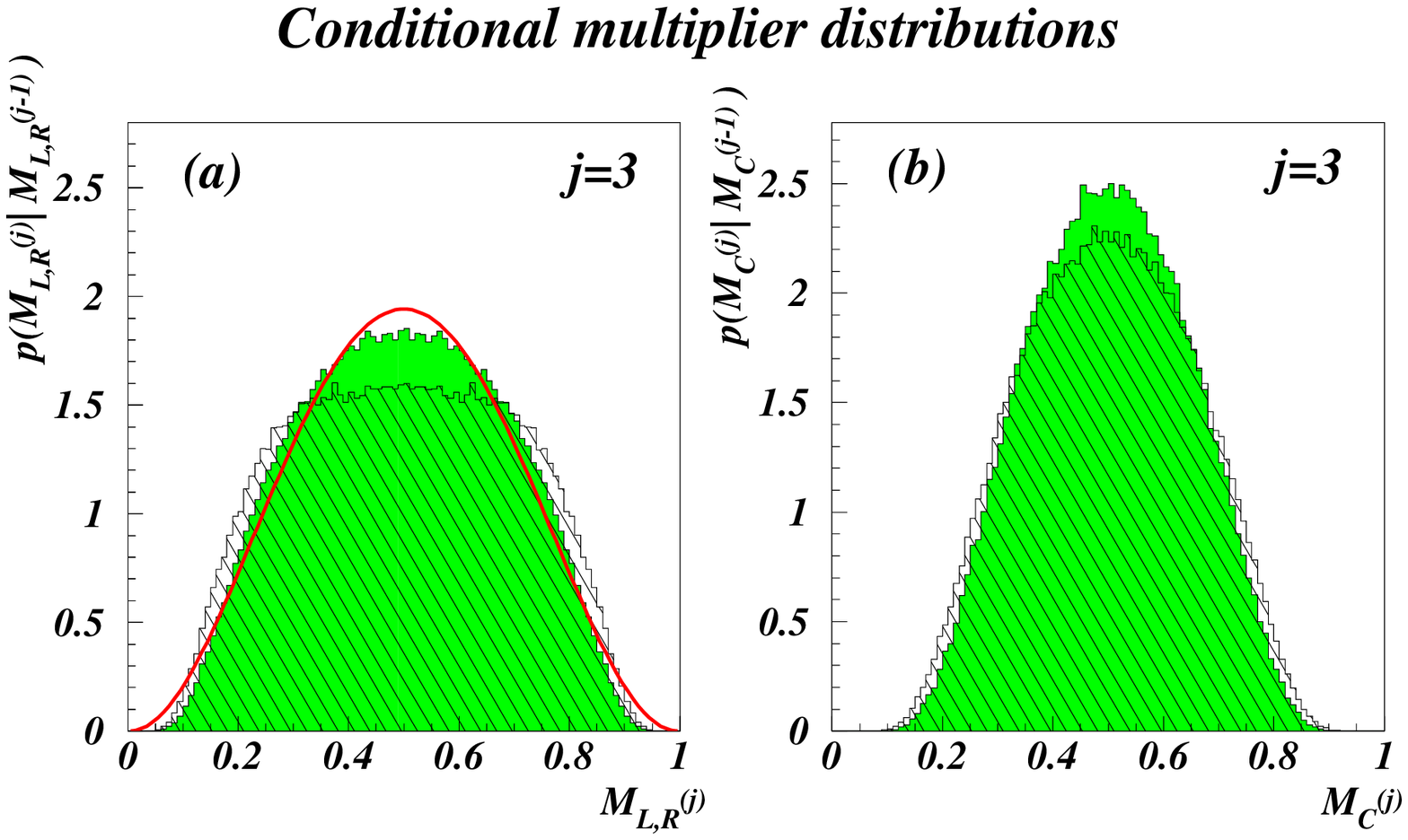,width=18cm}
\caption{
Conditional (a) left/right and (b) centred multiplier distributions for 
scale $j=3$ obtained from the splitting function ({\protect\ref{31eins}})
with $\alpha=0.4$ and $J=9$. 
(a) $p ( M_{L/R}^{(j)} | 
     \underline{M} \leq M_{L/R}^{(j-1)} \leq \overline{M} )$
for $\underline{M}=0.2$, $\overline{M}=0.4$ (grey) and 
$\underline{M}=0.6$, $\overline{M}=0.8$ (hatched);
(b) $p ( M_C^{(j)} | \underline{M} \leq M_C^{(j-1)} \leq \overline{M} )$
for $\underline{M}=0$, $\overline{M}=0.5$ (grey) and 
$\underline{M}=0.5$, $\overline{M}=1$ (hatched).
For comparison the experimentally deduced 
Beta-function parametrisation ({\protect\ref{22vier}}) of the left 
(right) unconditional multiplier distribution with 
$\beta=3.2$ is shown as a solid curve in (a).
} 
\end{centering}
\end{figure}

\newpage
\begin{figure}
\begin{centering}
\epsfig{file=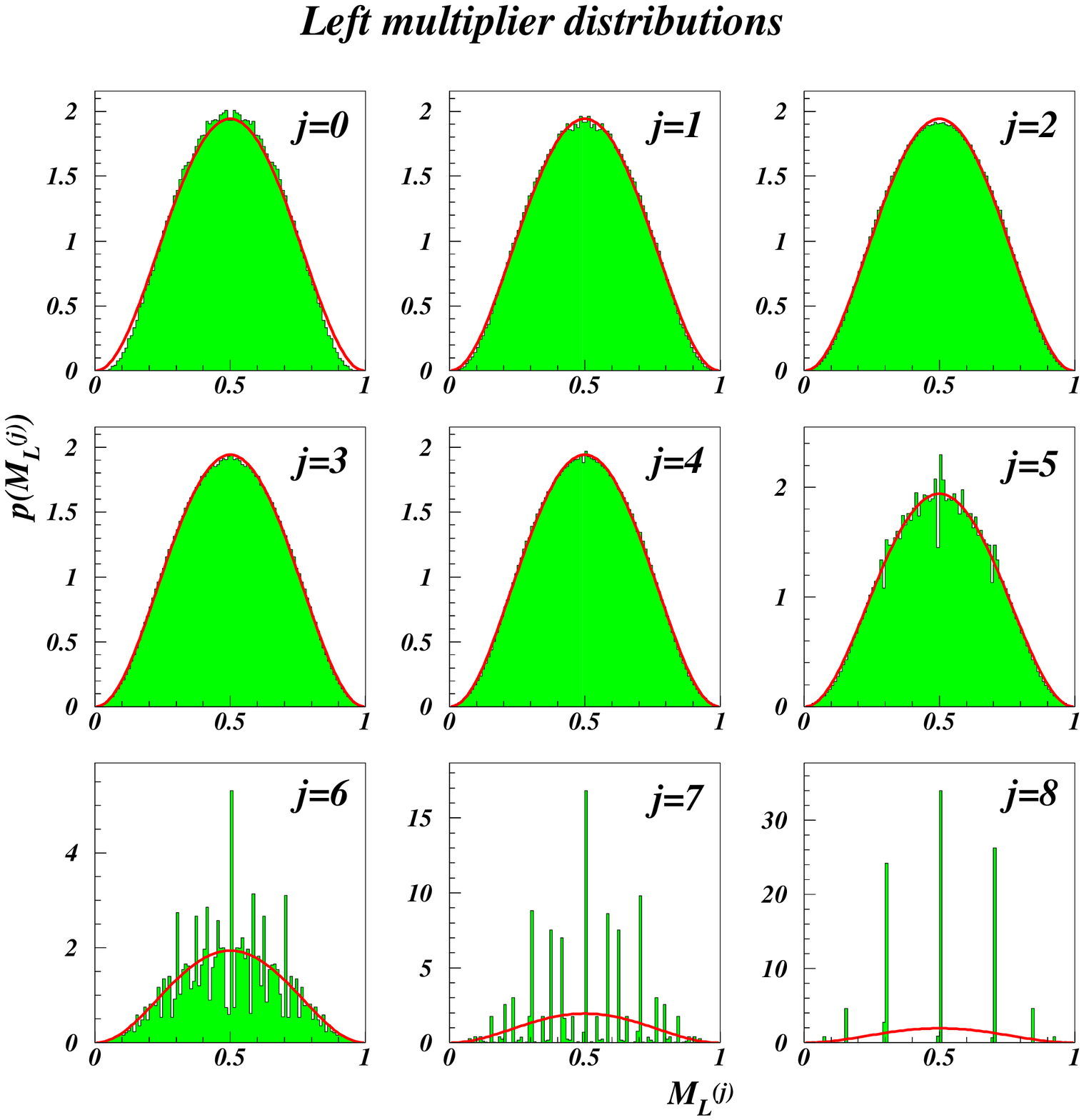,width=18cm}
\end{centering}
\end{figure}
\newpage
\begin{figure}
\begin{centering}
\epsfig{file=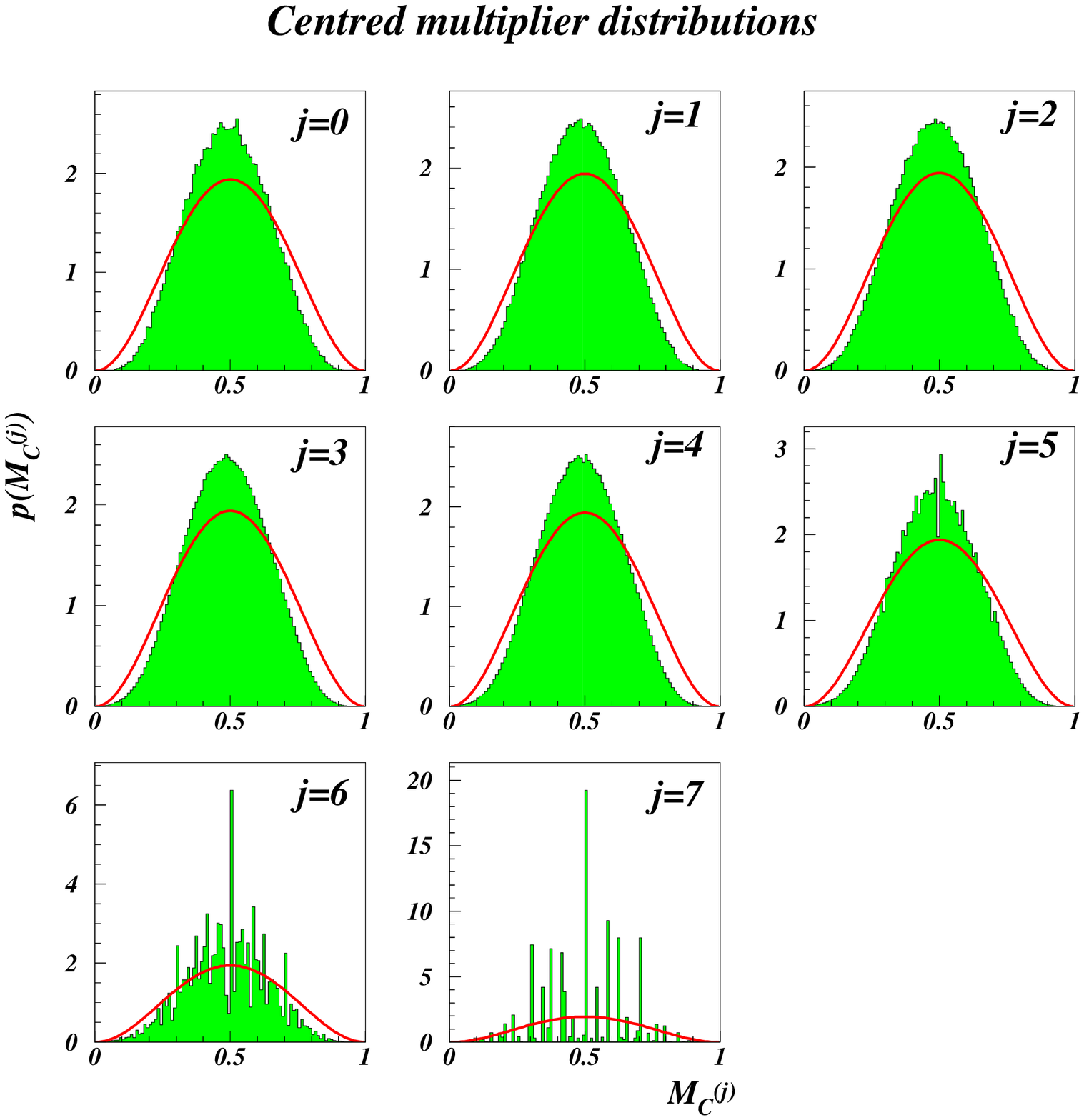,width=18cm}
\caption{
Same as Fig.\ 5, but now after restoration of translational invariance.
} 
\end{centering}
\end{figure}

\newpage
\begin{figure}
\begin{centering}
\epsfig{file=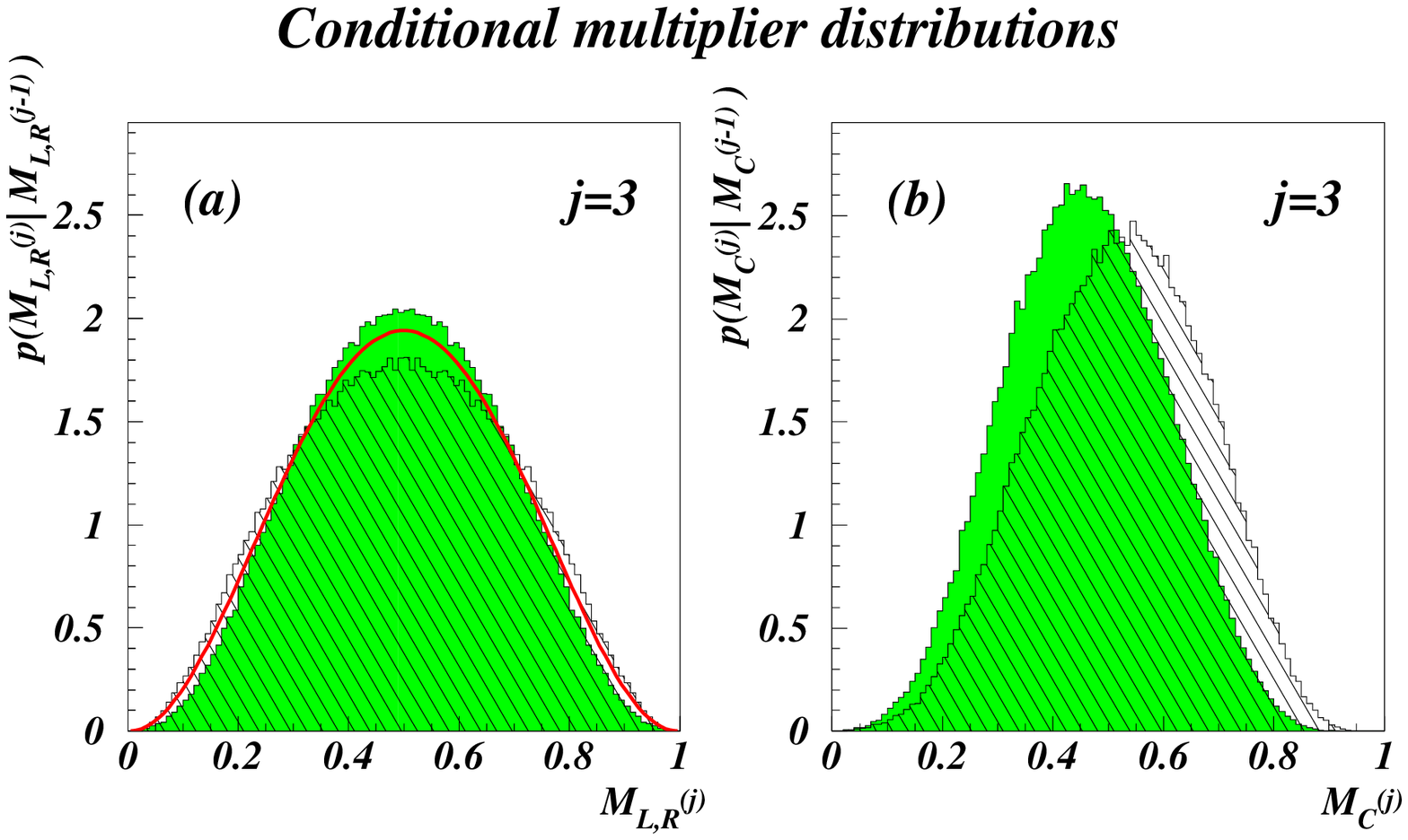,width=18cm}
\caption{
Same as Fig.\ 6, but now after restoration of translational invariance.
} 
\end{centering}
\end{figure}

\newpage
\begin{figure}
\begin{centering}
\epsfig{file=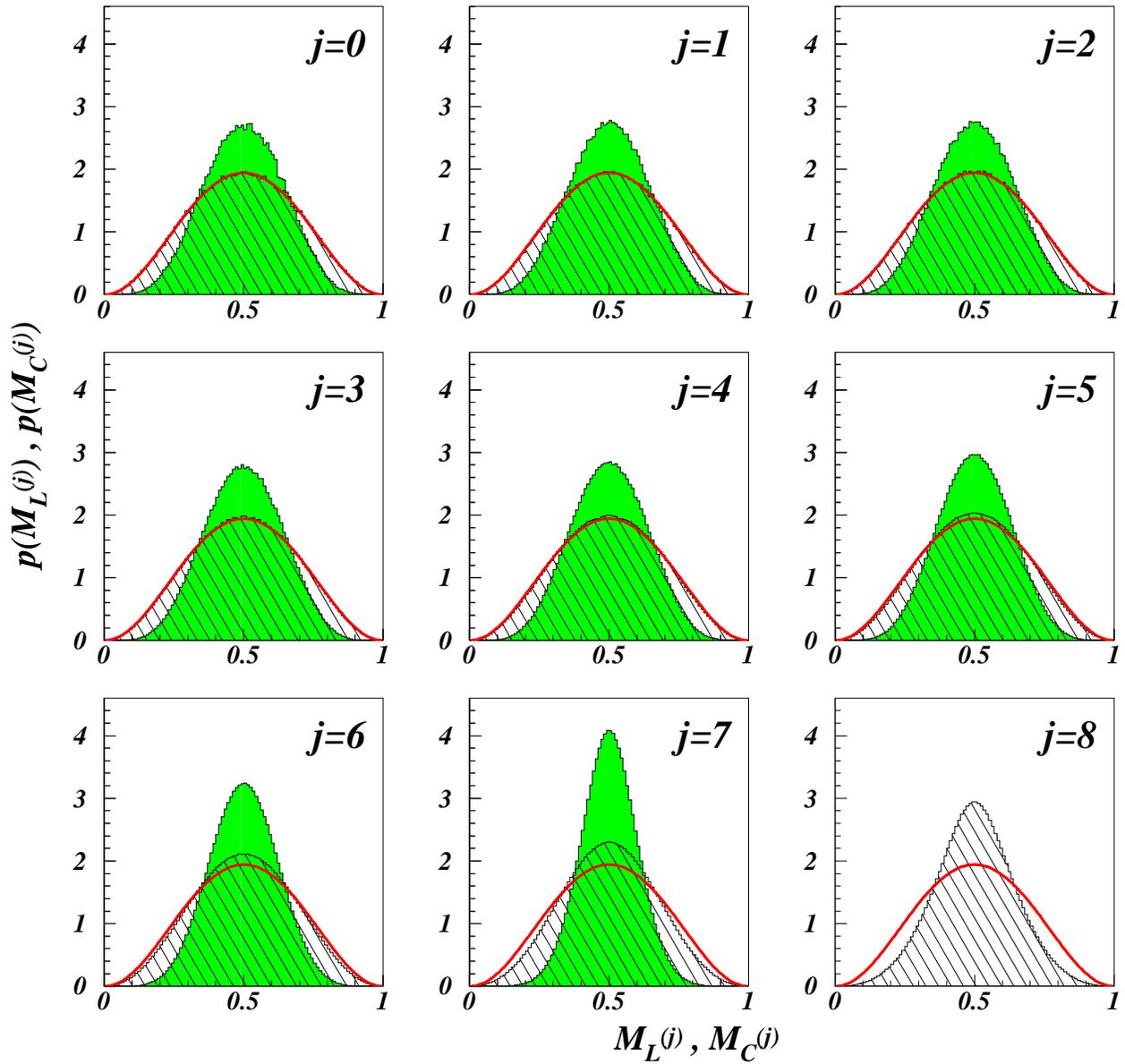,width=18cm}
\caption{
Convergence of the left (hatched) 
and central (grey) unconditional multiplier distributions 
$p(M_L^{(j)})$ and $p(M_C^{(j)})$ of the 
Beta-distribution cascade model
({\protect\ref{4zwei}}) to respective limiting forms.
Parameters are $J=9$, 
$\beta=3.2$. For comparison the experimentally deduced 
Beta-function parametrisation ({\protect\ref{22vier}}) 
of the left (right) unconditional multiplier distribution with 
$\beta=3.2$ is shown as a solid line. 
} 
\end{centering}
\end{figure}

\newpage
\begin{figure}
\begin{centering}
\epsfig{file=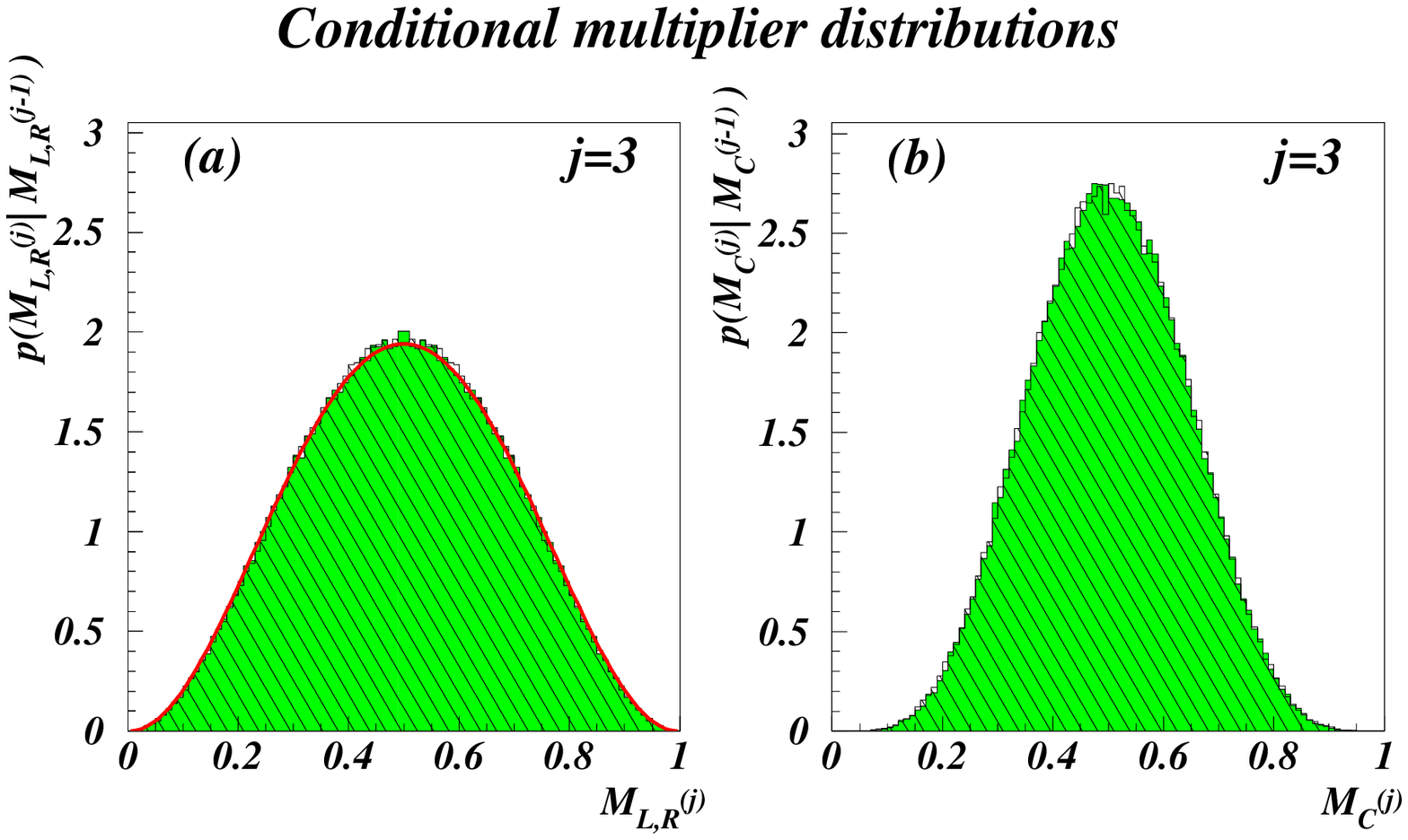,width=18cm}
\caption{
Conditional left (a) and centred (b) multiplier distributions for 
scale $j=3$ obtained from the splitting function ({\protect\ref{4zwei}})
with $\beta=3.2$ and $J=9$. 
(a) $p ( M_{L/R}^{(j)} | 
     \underline{M} \leq M_{L/R}^{(j-1)} \leq \overline{M} )$
for $\underline{M}=0.2$, $\overline{M}=0.4$ (grey) and 
$\underline{M}=0.6$, $\overline{M}=0.8$ (hatched);
(b) $p ( M_C^{(j)} | \underline{M} \leq M_C^{(j-1)} \leq \overline{M} )$
for $\underline{M}=0$, $\overline{M}=0.5$ (grey) and 
$\underline{M}=0.5$, $\overline{M}=1$ (hatched).
For comparison the experimentally deduced 
Beta-function parametrisation  ({\protect\ref{22vier}})
of the left (right) unconditional multiplier distribution with 
$\beta=3.2$ is shown as a solid curve in (a).
} 
\end{centering}
\end{figure}

\newpage
\begin{figure}
\begin{centering}
\epsfig{file=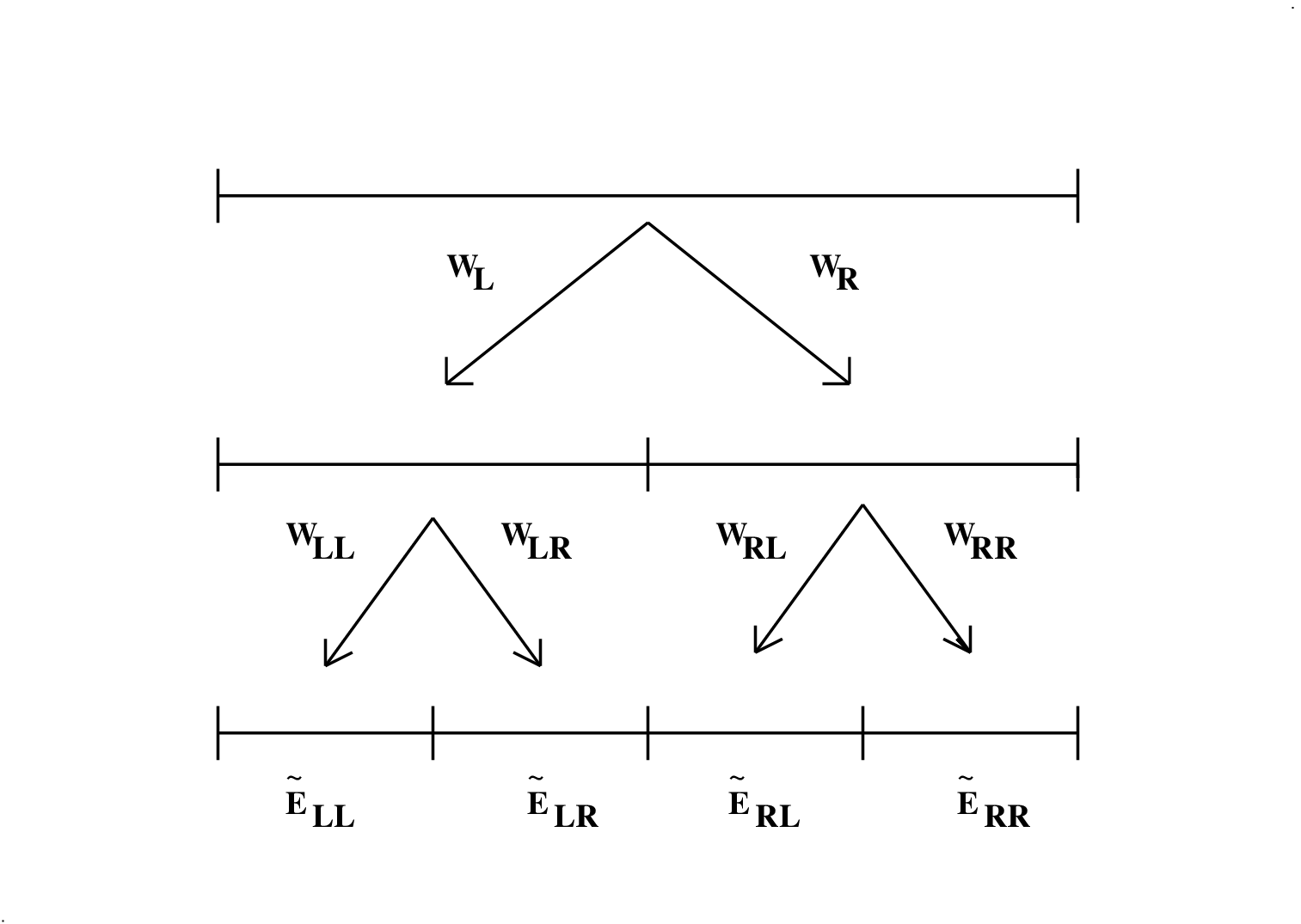,width=18cm}
\caption{
Weight curdling over two intermediate cascade steps.
} 
\end{centering}
\end{figure}

\newpage
\begin{figure}
\begin{centering}
\epsfig{file=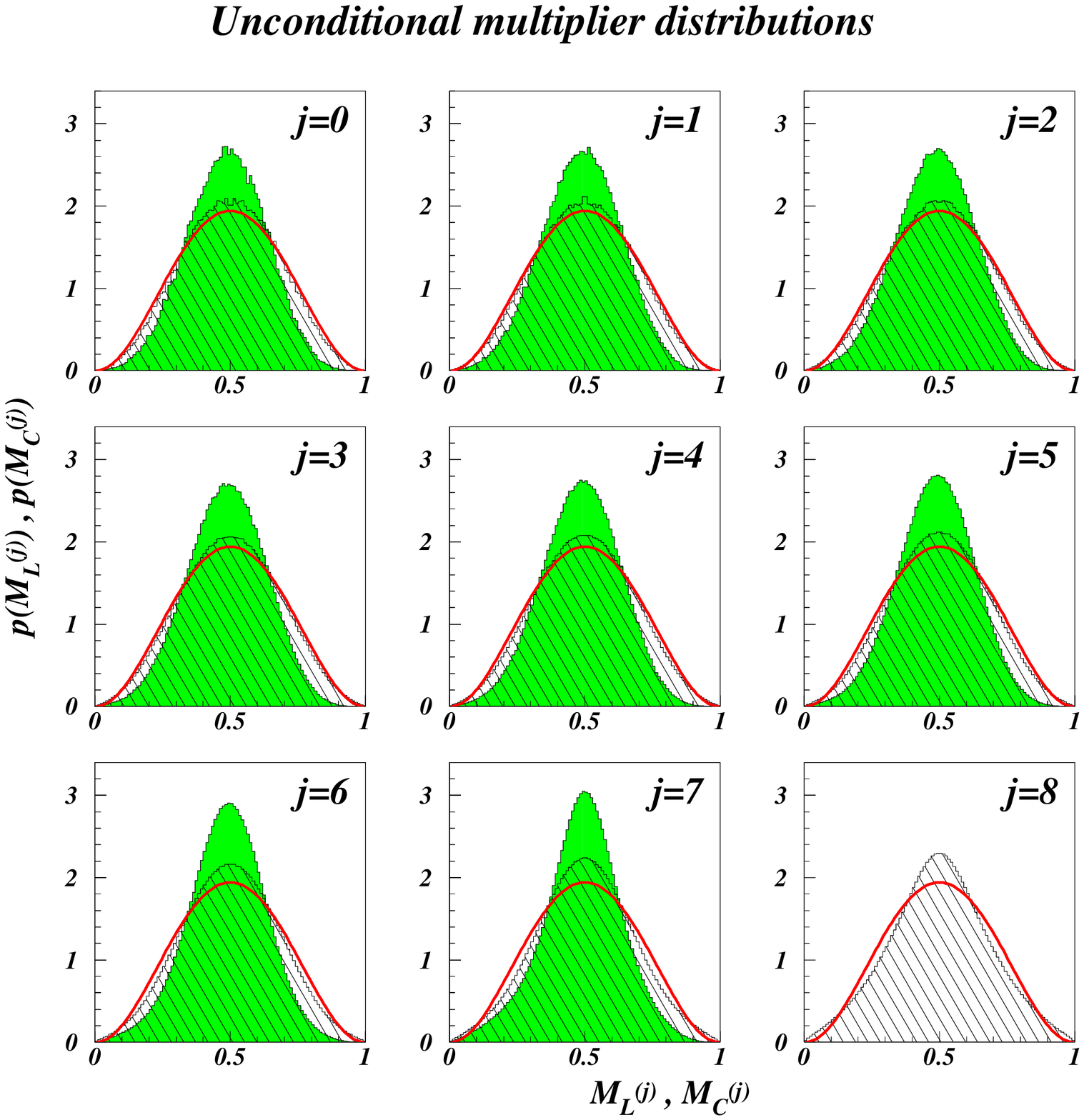,width=18cm}
\caption{
Same as Fig.\ 9, but now after restoration of translational invariance.
} 
\end{centering}
\end{figure}

\newpage
\begin{figure}
\begin{centering}
\epsfig{file=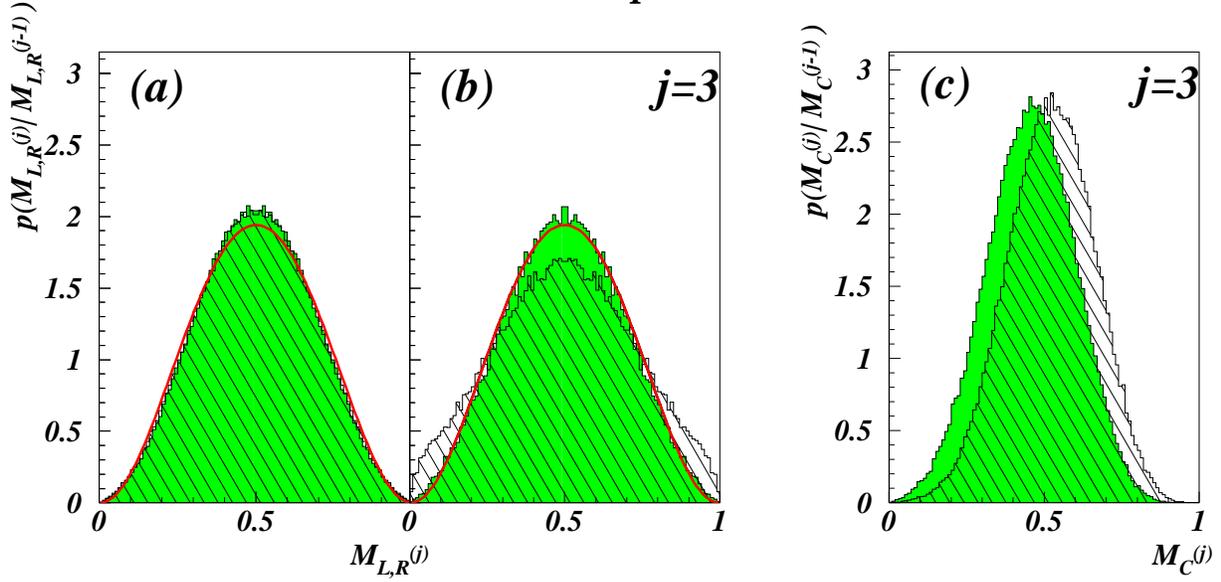,width=18cm}
\caption{
Conditional left (a,b) and centred (c) multiplier distributions for 
scale $j=3$ obtained from the splitting function ({\protect\ref{4zwei}})
with $\beta=3.2$ and $J=9$ after restoration of translational invariance. 
(a) $p ( M_{L/R}^{(j)} | 
     \underline{M} \leq M_{L/R}^{(j-1)} \leq \overline{M} )$
for $\underline{M}=0.2$, $\overline{M}=0.4$ (grey) and 
$\underline{M}=0.6$, $\overline{M}=0.8$ (hatched);
(b) $p ( M_{L/R}^{(j)} | 
     \underline{M} \leq M_{L/R}^{(j-1)} \leq \overline{M} )$
for $\underline{M}=0$, $\overline{M}=0.2$ (grey) and 
$\underline{M}=0.8$, $\overline{M}=1$ (hatched);
(c) $p ( M_C^{(j)} | \underline{M} \leq M_C^{(j-1)} \leq \overline{M} )$
for $\underline{M}=0$, $\overline{M}=0.5$ (grey) and 
$\underline{M}=0.5$, $\overline{M}=1$ (hatched).
For comparison the experimentally deduced 
Beta-function parametrisation  ({\protect\ref{22vier}})
of the left (right) unconditional multiplier distribution with 
$\beta=3.2$ is shown as a solid curve in (a,b).
} 
\end{centering}
\end{figure}

\newpage
\begin{figure}
\begin{centering}
\epsfig{file=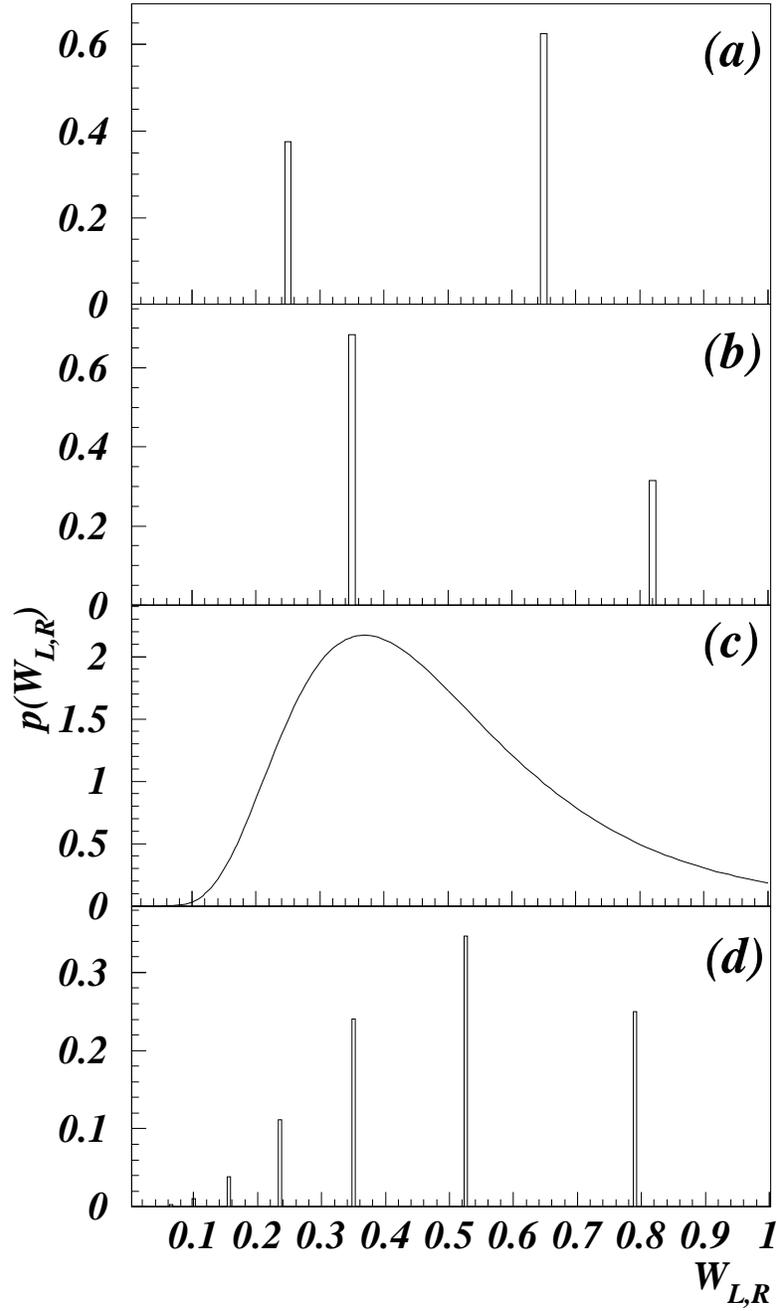,width=12cm}
\caption{
Asymmetric weight distributions:
(a) asymmetric binomial ({\protect\ref{51eins}}) with 
    $\alpha_1=0.5$, $\alpha_2=0.3$,
(b) asymmetric binomial ({\protect\ref{51eins}}) with 
    $\alpha_1=0.3$, $\alpha_2=0.65$,
(c) log-normal ({\protect\ref{52eins}}) with $\sigma^2=0.45$,
(d) log-Poisson ({\protect\ref{53eins}}) with $c_1=3c_2=2$.
} 
\end{centering}
\end{figure}

\newpage
\begin{figure}
\begin{centering}
\epsfig{file=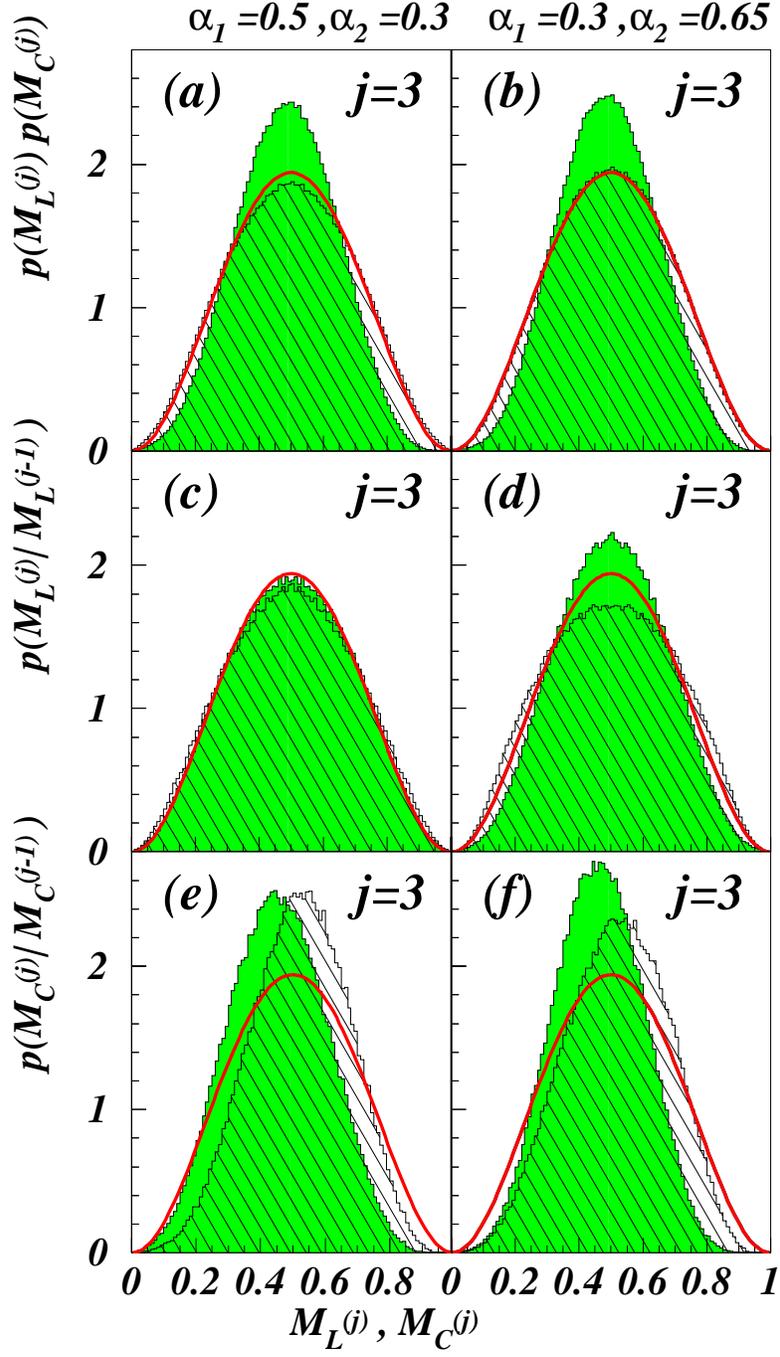,width=12cm}
\caption{
Various multiplier distributions resulting from the asymmetric
binomial model after restoration of translational invariance. 
Left column is for model parameters
$\alpha_1=0.5$, $\alpha_2=0.3$, $J=9$, and right column is for
$\alpha_1=0.3$, $\alpha_2=0.65$, $J=9$.
First line: unconditional left and centred
multiplier distributions $p(M_L)$ and $p(M_C)$ at $j=3$;
second line: conditional left/right multiplier distribution 
$p ( M_{L/R}^{(j)} | 
     \underline{M} \leq M_{L/R}^{(j-1)} \leq \overline{M} )$
for $\underline{M}=0.2$, $\overline{M}=0.4$ (grey) and 
$\underline{M}=0.6$, $\overline{M}=0.8$ (hatched) at $j=3$;
third line: conditional centred multiplier distribution
$p ( M_C^{(j)} | \underline{M} \leq M_C^{(j-1)} \leq \overline{M} )$
for $\underline{M}=0$, $\overline{M}=0.5$ (grey) and 
$\underline{M}=0.5$, $\overline{M}=1$ (hatched) at $j=3$.
For comparison the experimentally deduced 
Beta-function parametrisation ({\protect\ref{22vier}})
of the left (right) unconditional multiplier distribution with 
$\beta=3.2$ is shown as a solid curve 
in each plot.
} 
\end{centering}
\end{figure}

\newpage
\begin{figure}
\begin{centering}
\epsfig{file=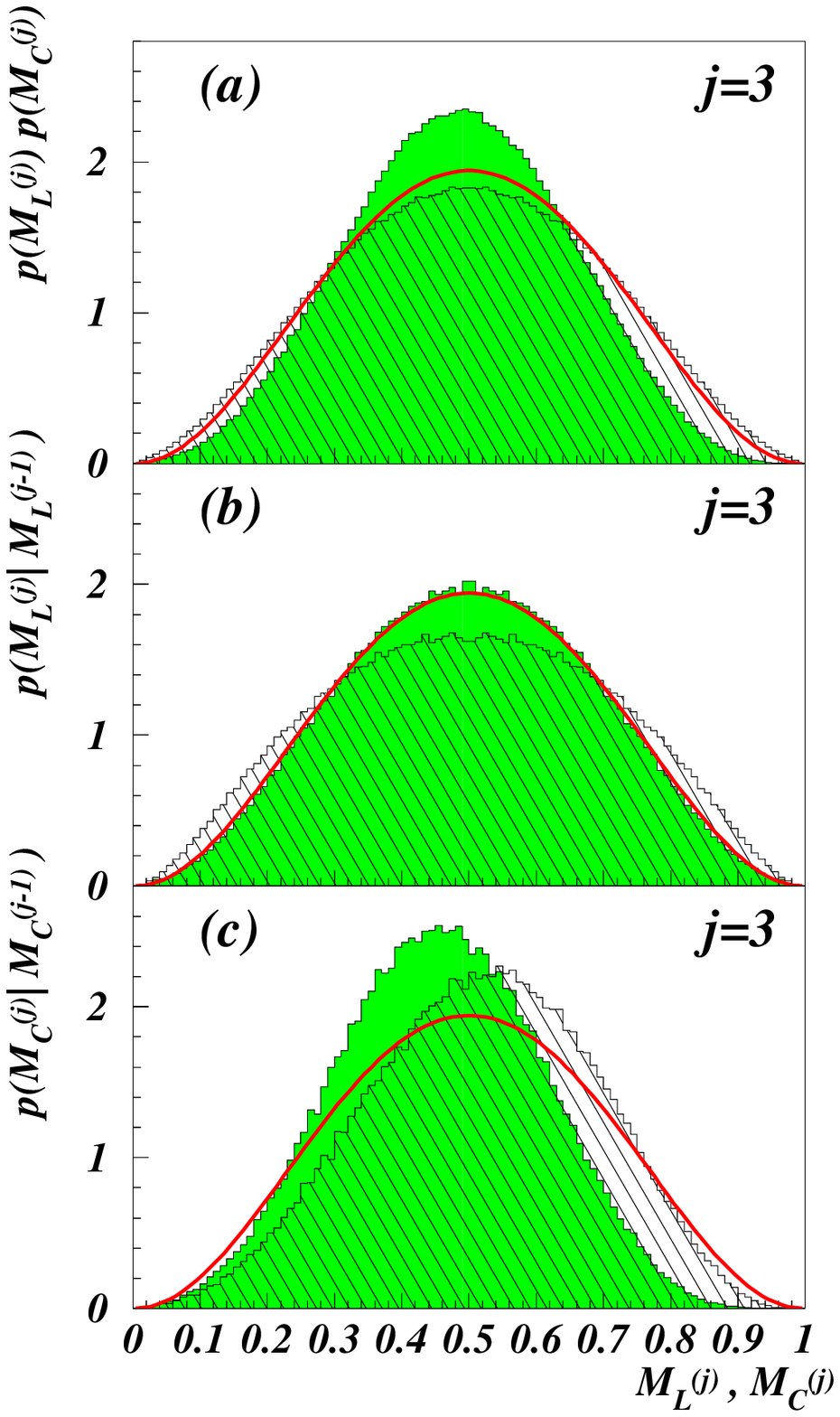,width=12cm}
\caption{
Various multiplier distributions resulting from the 
log-normal-model after restoration of translational invariance;
parameters are $\sigma^2=0.45$, $J=9$.
(a) unconditional left and centred
multiplier distributions $p(M_L)$ and $p(M_C)$ at $j=3$;
(b) conditional left/right multiplier distribution 
$p ( M_{L/R}^{(j)} | 
     \underline{M} \leq M_{L/R}^{(j-1)} \leq \overline{M} )$
for $\underline{M}=0.2$, $\overline{M}=0.4$ (grey) and 
$\underline{M}=0.6$, $\overline{M}=0.8$ (hatched) at $j=3$;
(c) conditional centred multiplier distribution
$p ( M_C^{(j)} | \underline{M} \leq M_C^{(j-1)} \leq \overline{M} )$
for $\underline{M}=0$, $\overline{M}=0.5$ (grey) and 
$\underline{M}=0.5$, $\overline{M}=1$ (hatched) at $j=3$.
For comparison the experimentally deduced 
Beta-function parametrisation ({\protect\ref{22vier}})
of the left (right) unconditional multiplier distribution with 
$\beta=3.2$ is shown as a solid curve in each plot.
} 
\end{centering}
\end{figure}

\newpage
\begin{figure}
\begin{centering}
\epsfig{file=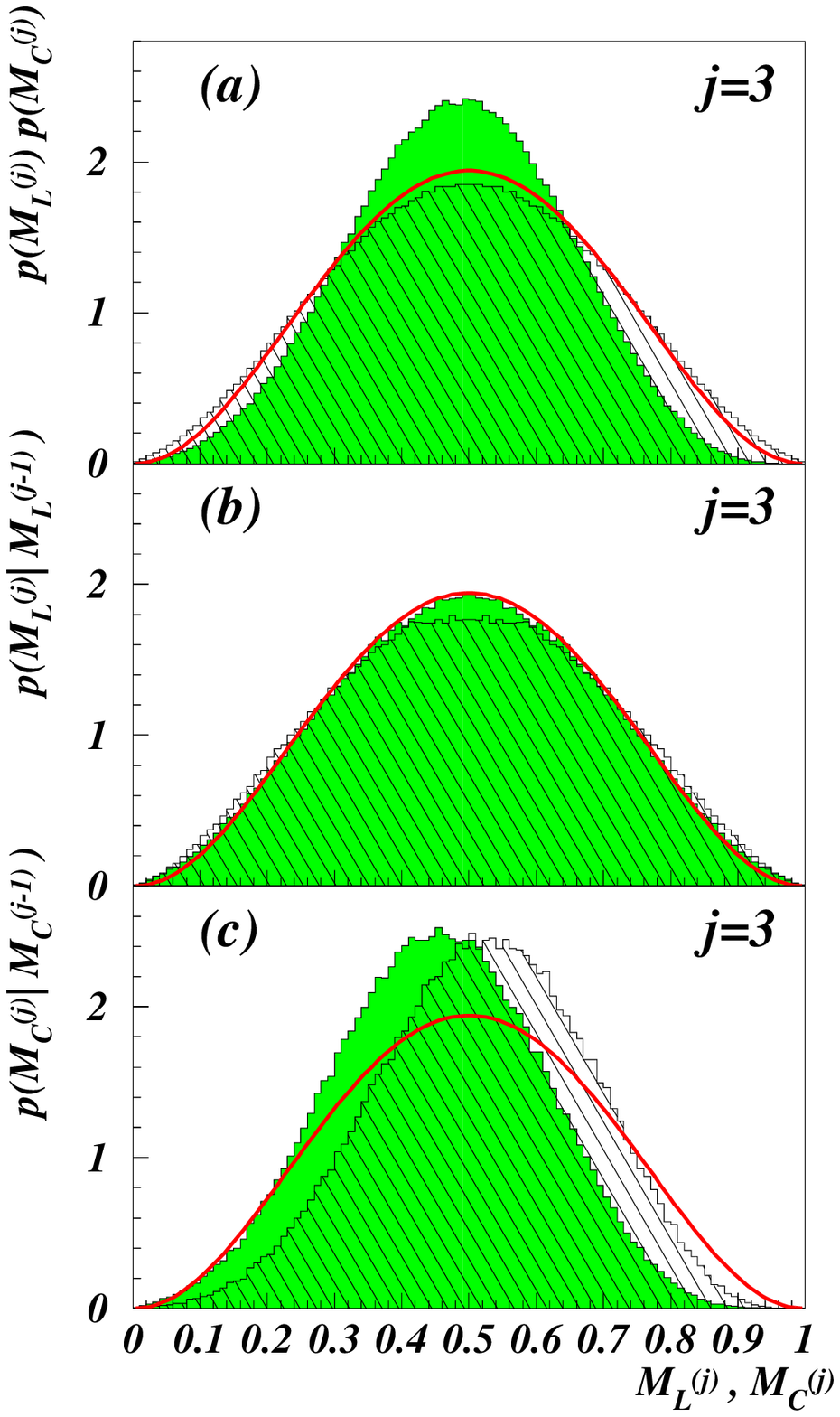,width=12cm}
\caption{
Various multiplier distributions resulting from the 
log-Poisson-model after restoration of translational invariance;
parameters are $c_1=3c_2=2$, $J=9$.
(a) unconditional left and centred
multiplier distributions $p(M_L)$ and $p(M_C)$ at $j=3$;
(b) conditional left/right multiplier distribution 
$p ( M_{L/R}^{(j)} | 
     \underline{M} \leq M_{L/R}^{(j-1)} \leq \overline{M} )$
for $\underline{M}=0.2$, $\overline{M}=0.4$ (grey) and 
$\underline{M}=0.6$, $\overline{M}=0.8$ (hatched) at $j=3$;
(c) conditional centred multiplier distribution
$p ( M_C^{(j)} | \underline{M} \leq M_C^{(j-1)} \leq \overline{M} )$
for $\underline{M}=0$, $\overline{M}=0.5$ (grey) and 
$\underline{M}=0.5$, $\overline{M}=1$ (hatched) at $j=3$.
For comparison the experimentally deduced 
Beta-function parametrisation ({\protect\ref{22vier}}) 
of the left (right) unconditional multiplier distribution with 
$\beta=3.2$ is shown as a solid curve in each plot.
} 
\end{centering}
\end{figure}

\end{document}